\documentclass[11pt,eqsecnum]{article}
\usepackage{amsmath,amsfonts,amssymb,latexsym,theorem,mathrsfs,latexsym,cite,setspace,
comment,color,multirow,graphicx}
\usepackage{hyperref}
\pdfoutput=1

\newcommand{\ma}[1]{\mbox{$\mathcal{#1}$}}

\newcommand{\D}{{\rm d}}
\newcommand{\ti}{\tilde}

\begin{document}

\begin{titlepage}

\begin{flushright}
{
\small 
YITP-20-133
\\
\today
}
\end{flushright}
\vspace{1cm}

\begin{center}
%---------- title ----------%
{\LARGE \bf
\begin{spacing}{1}
Static spacetimes haunted by a phantom scalar field II: 
\\ \vspace{0.2cm}
dilatonic charged solutions
\end{spacing}
}
\end{center}
\vspace{.5cm}

\begin{center}
%---------- author ----------%
{\large \bf
Masato Nozawa
} \\

\vskip 1cm
{\it
Center for Gravitational Physics, Yukawa Institute for Theoretical Physics, \\
 Kyoto University, Kyoto 606-8502, Japan.}\\
\texttt{masato.nozawa@yukawa.kyoto-u.ac.jp}

\end{center}

\vspace{.5cm}

%---------- abstract ----------%
\begin{abstract}
We present a method to generate static solutions in the Einstein-Maxwell system with a (phantom) dilaton field in $n(\ge 4)$-dimensions, based upon the symmetry of the target space for the nonlinear sigma model. Unlike the conventional Einstein-Maxwell-dilaton system, there appears a critical value of the coupling constant for a phantom dilaton field. In the noncritical case, the target space is $\mathbb R\times {\rm SL}(2,\mathbb R)/H$ with the maximal subgroup $H=\{{\rm SO}(2), {\rm SO}(1,1)\}$, whereas in the critical case the target space becomes a symmetric pp-wave and the corresponding Killing vectors form a non-semisimple algebra.  In either case, we apply the formalism to charge up the neutral solutions and show the analytical expression for dilatonic charged versions of (i) the Fisher solution, (ii) the Gibbons solution, and (iii) the Ellis-Bronnikov solution. 
We discuss global structures of these solutions in detail.
It turns out that some solutions contained in the Fisher and Gibbons classes possess the parallelly propagated (p.p) curvature singularities in the parameter region where all the scalar curvature invariants remain bounded. These p.p curvature singularities are not veiled by a horizon, thrusting them into physically untenable nakedly singular spacetimes. We also demonstrate that the dilatonic-charged Ellis-Bronnikov solution admits a parameter range under which the solution represents a regular  wormhole spacetime in the two-sided asymptotically flat regions.  
\end{abstract}

\vspace{.5cm}

\setcounter{footnote}{0}

\end{titlepage}

\setcounter{tocdepth}{2}
\tableofcontents

\newpage

%==============================================%
%<<<<<<<<<<<< SECTION I Introduction >>>>>>>>>>>>>>%
%==============================================%
\section{Introduction}

It seems fairly reasonable to impose standard energy conditions on the classical matter fields of physical interest~\cite{Hawking:1973uf,Maeda:2018hqu}, since their violation would put a curse on the spacetime to be unstable. Various monumental theorems in general relativity have been successfully proven together with energy conditions. This includes the singularity theorems~\cite{Penrose:1964wq,Hawking:1969sw}, the topological censorship \cite{Friedman:1993ty,Galloway}, the positive energy theorem~\cite{Schon:1979rg,SchonYau,Witten:1981mf} and the topology and area increasing theorem of black holes~\cite{Hawking:1971vc}.  Despite the physical soundness for the positivity of local energy density and the mathematical robustness of these theorems, the observation data for the current acceleration of the universe does not entirely exclude the possibility of phantom equations of state \cite{Aghanim:2018eyx}, which violates the null energy condition. It thus deserves a further attention to investigate the role of phantoms within classical gravitational theory.

A phantom scalar field--a scalar field with a reversed sign of the kinetic term--is the simplest exotic matter field that violates the energy conditions and has been 
actively investigated in the context of cosmology as a candidate of the dark energy~\cite{Caldwell:1999ew,Caldwell:2003vq,Dabrowski:2003jm,Nojiri:2003vn,Elizalde:2004mq}. 
Our interest here is devoted to the static spacetimes with a phantom scalar field. 
Static and spherically symmetric solutions with a conventional scalar field in $n=4$ dimensions are uniquely determined to be the solution originally constructed by Fisher~\cite{Fisher:1948yn}, which has been rediscovered repetitively by  Bergmann and Leipnik \cite{Bergmann:1957zza}, Buchdahl \cite{Buchdahl:1959nk},  Janis, Newman and Winicour \cite{jnw1968}, Ellis \cite{Ellis1973}, Bronnikov \cite{Bronnikov1973} and Wyman \cite{Wyman:1981bd}.  
In contrast, static and spherically symmetric spacetimes with a phantom scalar field admit two more distinct families of solutions on top of the phantom Fisher solution, as first demonstrated by Ellis in \cite{Ellis1973}. One is the ``exponential metric'' which was later rediscovered and refined by Gibbons~\cite{Gibbons:2003yj,Gibbons:2017jzk}, so that we shall refer to this solution as the Ellis-Gibbons solution. The gravitational attractive force and the repulsive force by the scalar field are strictly balanced for the Ellis-Gibbons solution, allowing one to obtain a multi-center solution like the Majumdar-Papapetrou solution in the Einstein-Maxwell system~\cite{Majumdar:1947eu,Papaetrou:1947ib}.
The third solution has been also found independently by Bronnikov \cite{Bronnikov1973}, which we call the Ellis-Bronnikov solution. The Ellis-Bronnikov solution describes a wormhole connecting two asymptotically flat regions.

Wormholes epitomize the spacetime in which the the energy conditions are violated and represent the bridge structure of spacetimes connecting two disjoint universes. 
Since the seminal paper by Ellis \cite{Ellis1973} in which the first eponymous model of wormholes was put forward,  a number of works have been done concerning the global structure \cite{Morris:1988cz,Hochberg:1998ha,Kim:2013tsa}, stability \cite{Bronnikov:2013coa,Cremona:2018wkj} and possible time travels \cite{Morris:1988tu} in wormhole spacetimes. We refer the reader to see~\cite{visser,Lobo:2007zb} for a comprehensive review and a complete list of references.

Also, wormhole geometries are of help for the probe of the extra dimensions.  
Wormholes in higher dimensions have been studied by many authors \cite{Clement:1983fe,Bronnikov:1997gj,Torii:2013xba}, but the extension of Ellis' work has not been addressed until recently. In our previous paper \cite{paperI}, we have performed a complete classification of the static solutions with pseudo-spherical symmetry sourced by a phantom scalar field and elucidated their  global structure. 
In the past studies, most attention has been restricted exclusively to the scalar curvature singularities. 
Our important discovery in \cite{paperI} is that the Fisher and the Ellis-Gibbons solutions in arbitrary dimensions necessarily develop a naked p.p curvature singularity in the parameter region where there exist no scalar curvature singularities. This type of singularity is characterized by the divergence of the Riemann tensor component in a frame that is parallelly propagated along some curve \cite{Hawking:1973uf}.\footnote{For the present analysis these curves are taken to be radial null geodesics.} 
We therefore concluded that these solutions fail to describe regular wormholes, in contrast to some claims in the literature. The maximal extension of the Ellis-Bronnikov spacetime is fairly delicate in higher dimensions, but has been successfully done, allowing us to physically interpret  this solution as a genuine wormhole.

In the context of string theory and supergravity, the coupling of a scalar field with the higher-rank gauge fields is ubiquitous. A primitive system of this sort is the Einstein-Maxwell-dilaton gravity with the exponential coupling $e^{-2a\phi}$. 
An extensive study of static and spherically symmetric solutions revealed that properties of a black hole in four dimensions are rather sensitive to the dilaton coupling for a normal~\cite{Gibbons:1987ps,Garfinkle:1990qj} and for a 
phantom scalar field \cite{Gibbons:1996pd}. 
In the phantom case, there appear  black holes with unusual causal structures~\cite{Clement:2009ai,AzregAinou:2011rj}. 
 In order to gain more insights into the effect of the electric charge, 
 it is fully encouraging  to obtain explicit form of exact solutions and their exhaustive list in 
 $n$ dimensions. However, 
 the field equations become substantially unwieldy in the presence of dilaton coupling and in higher dimensions. This is indeed the case even in the absence of the Maxwell field in higher dimensions, 
 where the exact wormhole solutions are not obtained in a closed form unless a suitable choice of radial coordinate is made \cite{Torii:2013xba}.  An attempt to solve Einstein's equations directly is not therefore a practical expedient.  

 A powerful method to overcome this drawback is the solution-generating technique, by utilizing the underlying symmetry of the action (see e.g \cite{Clement:2008qx,Galtsov:2008zz}).
In the present paper, we establish a method to generate static solutions in the Einstein-Maxwell-(phantom-)dilaton system in arbitrary $n(\ge 4)$ dimensions, for which the system can be reduced to the $(n-1)$-dimensional gravity coupled to the nonlinear sigma model. 
The solution-generating method in the Einstein-Maxwell-dilaton gravity has been discussed 
in~\cite{Galtsov:1995mb,Yazadjiev:2005hr} for a conventional scalar field 
and in~\cite{Gibbons:1996pd,AzregAinou:2011rj} for a phantom scalar field in four dimensions.   Central to our result is the existence of a critical value of the dilaton coupling in the phantom case, at which the symmetry of the target space changes substantially. 
In~\cite{Gibbons:1996pd}, the authors have noticed that the separate analysis is demanded at this coupling in four dimensions, but it has been left open why this exceptional value occurs. Our treatment in the symmetry algebra from the sigma model perspective clearly answers this question. 

As applications of our methodology, we construct the dilatonic charged versions of the Fisher, the Gibbons, and the Ellis-Bronnikov solutions, in line with whether the coupling is critical or not. Our primary motivation is to see whether the electric charge alters the causal structure of each spacetime.
We find that the charged Ellis-Bronnikov solution describes a regular charged wormhole in two-sided asymptotically flat regions under a certain parameter regime. 
Last but not least we wish to point out that some ``black holes'' contained in a family of charged Fisher class admit a p.p curvature singularity at the horizon, while all the scalar curvature invariants remain finite there. In the literature, these solutions were argued to be black holes which are regular on and outside the event horizon. Our argument shows that this conclusion is dubious by the presence of a p.p curvature singularity.

The structure of the present paper is as follows. 
In the following section, we give a preliminary discussion for the Einstein-Maxwell-(phantom-)dilaton system. 
In section \ref{sec:noncritical}, we 
derive finite transformations to generate new static solutions from neutral ones for the noncritical case.
The critical case is explored in section \ref{sec:critical}. 
We detail the construction of the dilatonic charged versions of the Fisher, Gibbons, and Ellis-Bronnikov solutions in these sections. The conditions under which each solution deserves a black hole/wormhole will be clarified. Concluding remarks and future prospects are given in the final section \ref{sec:conclusion}.
An appendix provides supplementary material for curvature tensors.

Our basic notations follow~\cite{wald}.
The conventions of curvature tensors are 
$[\nabla _\rho ,\nabla_\sigma]V^\mu ={R^\mu }_{\nu\rho\sigma}V^\nu$ 
and ${R}_{\mu \nu }={R^\rho }_{\mu \rho \nu }$.
The Lorentzian metric is taken to be the mostly plus sign, and 
Greek indices run over all spacetime indices. 
%We denote the $n$-dimensional gravitational constant to be $\kappa_n=8\pi G_n$ for brevity.

%==================================================%
%<<<<<<<<<<<< SECTION II charged solutions>>>>>>>>>>>>>>%
%==================================================%
\section{Einstein-Maxwell-dilaton system}
\label{sec:setup}

In this paper, 
we consider the $n(\ge 4)$-dimensional Einstein-Maxwell-(phantom-)dilaton system described by the following action
\begin{align} 
S_{(n)}&=\frac{1}{16\pi G_n}\int \D^nx\sqrt{-g}\biggl({R}-2\epsilon(\nabla \varphi)^2-e^{-2a\varphi}F_{\mu\nu}F^{\mu\nu}\biggl)\,. \label{action-dil2}
\end{align}
where $F=\D A$ is the field strength of the Abelian gauge field and $a$ is the dilaton coupling constant. 
$\epsilon=+1$ for an ordinary scalar field, whereas $\epsilon=-1$ for a phantom field. 
The (phantom) scalar field is not canonically normalized, but this normalization is conventional 
(see e.g., \cite{Gibbons:1996pd,Gibbons:1994vm,Galtsov:1995mb,{Clement:2009ai},Nozawa:2010rf,Nozawa:2018kfk}) and makes contact with the literature in the context of the Kaluza-Klein paradigm.

The field equations following from the action (\ref{action-dil2}) are given by 
\begin{subequations}
\begin{align}
{R}_{\mu\nu}-\frac12g_{\mu\nu}{R}&=T^{(\rm em)}_{\mu\nu} +T^{(\varphi)}_{\mu\nu}\,, \label{efe} \\
\nabla_\nu (e^{-2a\varphi}F^{\mu\nu})&=0\,,\\
 \epsilon\Box \varphi+\frac{a}{2}e^{-2a\varphi}F_{\rho\sigma}F^{\rho\sigma}& =0\,, \label{phi}
\end{align}
\end{subequations} 
where the energy-momentum tensors for the gauge field ${A}_{\mu}$ and a dilaton field $\varphi$ are given by 
\begin{subequations}
\begin{align}
T^{(\rm em)}_{\mu\nu}=&2e^{-2a\varphi}\biggl(F_{\mu\rho}F_{\nu}^{~\rho}-\frac{1}{4}g_{\mu\nu}F_{\rho\sigma}F^{\rho\sigma}\biggl)\,, \label{Tab-Max}\\
T^{(\varphi)}_{\mu\nu} =&2\epsilon\biggl((\nabla_\mu\varphi)(\nabla_\nu\varphi)-\frac12g_{\mu\nu}(\nabla\varphi)^2\biggl)\,.\label{Tab-scalar}
\end{align}
\end{subequations}

\subsection{Static truncation with a nonlinear sigma model}

Let us focus on the $n(\ge 4)$-dimensional spacetimes admitting a hypersurface-orthogonal timelike Killing vector $\xi$. Without losing any generality, the metric in this class of spacetimes can be cast into
\begin{align}
\D s^2=&- e^{-2a_{c}\sigma}\D t^2+e^{2a_{c}\sigma/(n-3)}{g}_{IJ}\D y^I \D y^J, \label{metric}
\end{align}
where $t=x^0$, $y^I~(I=1,2,\cdots,n-1)$ are the coordinates on the Riemannian base space ($M^{n-1}$, $g_{IJ}$) and 
\begin{align}
\label{ac}
a_{c}\equiv \sqrt{\frac{2(n-3)}{n-2}}.
\end{align}
${\sigma}$ is pertinent to the norm of a timelike Killing vector $\xi=\partial/\partial t$, implying that 
$\sigma$ and $g_{IJ}$ are independent of $t$.

We  assume that the electromagnetic field and the dilaton field are also static
$\mathcal L_\xi F_{\mu\nu}=\mathcal L_\xi \varphi=0$, and that the electromagnetic field
is purely electric $F_{IJ}=0$. The Bianchi identity $\D F=0$ 
alludes the existence of an electrostatic potential $E$ (modulo addition of a constant) such that $F_{tI}=-\partial_I E$ with $\mathcal L_\xi E=0$. 
Following the standard dimensional reduction of the Kaluza-Klein paradigm, 
we obtain the $(n-1)$-dimensional system of gravity coupled to a nonlinear sigma model 
$\phi^{a}=\{\sigma, \varphi, E\}~(a=1,2,3)$:
\begin{align}
\label{harmonicmap}
S_{(n-1)}=&\frac 1{2} \int \D^{n-1}y\sqrt{\det({g}_{IJ})}\biggl({}^{(n-1)}{R}-{G}_{ab}(\phi)g^{IJ}(D_I\phi^{a})(D_J\phi^{b})\biggl),
\end{align}
where $D_I$ is the covariant derivative on $M^{n-1}$ with respect to 
the metric ${g}_{IJ}$ of the base space, ${}^{(n-1)}{R}$ is the Ricci scalar constructed from $g_{IJ}$, and  $G_{ab}$ is the metric of the three-dimensional target space 
$\D s_T^2=G_{ab}\D\phi^{a}\D\phi^{b} $ with 
\begin{align}
\D s_T^2
=&2\Bigl(\D \sigma^2+\epsilon\D \varphi^2-e^{2a_{c}\sigma-2a\varphi}
\D E^2\Bigl).
\label{targetmet}
\end{align}
The equations of motion derived from the reduced action (\ref{harmonicmap}) are gravity-coupled harmonic maps \cite{Misner:1978am,Carter} with the following form
\begin{align}
\label{EOMNLSM1}
{}^{(n-1)}R_{IJ}={G}_{ab}(D_I\phi^{a})(D_J \phi^{b})\,,
\end{align}
and 
\begin{align}
\label{EOMNLSM2}
D^ID_I \phi^a+\Gamma^{a}{}_{bc}g^{IJ}(D_I\phi^{b})(D_J\phi^{c})=0 \,,
\end{align}
where $\Gamma^{a}{}_{bc}$ is the Levi-Civita connection composed of the target space metric $G_{ab}$.

In the subsequent sections, we exploit the symmetry of the target space 
(\ref{targetmet}) to generate various new solutions from an old one. 
The reasoning is substantiated as follows. 
Let us suppose that  the target space metric (\ref{targetmet}) admits a Killing vector
$\xi^a$ satisfying Killing's equations
$\mathcal D_a \xi_b+\mathcal D_b \xi_a=0$, where 
$\mathcal D_a$ is the linear connection of the target space.  It then follows that 
the infinitesimally transformed field $\phi^{\prime(a)}=\phi^{(a)}+\varepsilon \xi^a$ 
($\varepsilon \ll 1$)
also satisfies the harmonic map equation
\begin{align}
\label{}
&\left.D^ID_I(\phi^{a}+\varepsilon \xi^a)+\Gamma^{a}{}_{bc}
(\phi+\varepsilon \xi)D_I(\phi^{b}+\varepsilon \xi^b)D^I(\phi^{c}+\varepsilon \xi^c)\right|_{O(\varepsilon)}
\notag \\
&=  D_I \phi^{b}D^I \phi^{c} (\mathcal D_c \mathcal D_b \xi^a -\mathcal R
^a{}_{bcd}\xi^d)\notag \\
&=0\,.
\end{align}
Here we have posed the harmonic map equation 
(\ref{EOMNLSM2}) for $\phi^{a}$ and used the fact that any Killing vector $\xi^a$ must 
satisfy  $\mathcal D_c \mathcal D_b \xi^a -\mathcal R^a{}_{bcd}\xi^d=0$ (see e.g., \cite{wald}), 
where $\mathcal R^a{}_{bcd}$ is the Riemann tensor built out of the target space metric $G_{ab}$.
To be pedantic, this is nothing but the affine collineation, which maps a geodesic into another geodesic
(see section 35.4 in \cite{Stephani:2003tm}). A finite transformation is obtained by solving $\D \phi^{\prime a}/\D \varepsilon=\xi^a$
with initial conditions $\phi^{\prime a}|_{\varepsilon=0}=\phi^{a}$. 
Since the right-hand side of (\ref{EOMNLSM1}) remains also unaltered 
under the variation $\phi^{a}\to \phi^{\prime a}=\phi^{a}+\varepsilon \xi^a$, 
the new metric defined by $\phi^{\prime a}$ satisfies the 
equations of motion of the system. In this fashion, one can generate new solutions 
$\phi^{\prime a}$ from old $\phi^a$ without solving nonlinear PDEs, while preserving the 
base space metric $g_{IJ}$ invariant. 
This algebraic property is a testimony to the power of the solution-generating method.

Our impending problem is then to extract the symmetry algebra of the Killing vectors
for the target space (\ref{targetmet}), and to choose the appropriate seed solutions. 
As it turns out, the structure of the Killing symmetry relies heavily on 
whether $a_c^2+\epsilon a^2=0$ or not.  In what follows, we refer to the 
phantom case ($\epsilon=-1$) with $a=a_c$ as the {\it critical case}. 
We shall therefore divide the ensuing discussion if $a_c^2+\epsilon a^2=0$ or not.

\subsection{Neutral seed solutions}

Before proceeding into the charged case, 
let us first discuss the neutral case ($E=0$). The target space metric (\ref{targetmet}) 
is $\mathbb R^2$ for $\epsilon=+1$ and $\mathbb R^{1,1}$ for $ \epsilon=-1$. 
It is worth stressing that both of these flat target spaces are derivable from the ($n+1$)-dimensional metric satisfying {\it vacuum} Einstein's equations. 
In the $\epsilon=1$ (non-phantom) case, the embedding is given by 
\begin{align}
\label{}
\D s_{(n+1)}^2 = e^{-2\sqrt{(n-3)/(n-1)}\sigma}\left(-e^{2\varphi}\D t^2+e^{-2\varphi}\D w^2\right) 
+e^{4\sigma/\sqrt{(n-1)(n-3)}}g_{IJ}\D y^I \D y^J \,, 
\end{align}
where $t$, $w$ are two abelian Killing directions. In the above metric, 
the role of $\sigma$ and $\varphi$ can be interchanged and each sign of ($\sigma, \varphi$) 
is also irrelevant. 
For $\epsilon=-1$ case, we follow the argument in \cite{Gibbons:2017jzk} 
to find the desired embedding as 
\begin{align}
\label{}
\D s_{(n+1)}^2 = & e^{-2\sqrt{(n-3)/(n-1)}\sigma} \Bigl(-\cos(2\varphi)\D t^2+2 \sin(2\varphi)\D t\D w
+\cos(2\varphi)\D w^2\Bigr)\notag \\
&+e^{4\sigma/\sqrt{(n-1)(n-3)}}g_{IJ}\D y^I \D y^J \,.
\end{align}
In this case, one does not need to care about each sign of ($\sigma, \varphi$), but 
one cannot alternate the role of $\sigma$ and $\varphi$. 
%An interesting configuration for the $\epsilon=-1$ case is $\sigma=\pm \varphi$, corresponding to the null direction of the target space $G_{AB}\partial_I \phi^A \partial_J \phi^B=0$. 

As described above, the Killing symmetry of the target space is used to generate new solutions. 
In particular, we will show below that the charged solutions are obtainable from the neutral solutions. 
As neutral seed solutions of physical interest, 
we direct our primary attention to static metrics with the following form
\begin{align}
\label{pseudospher}
\D s^2  =-f_1(r)\D t^2+f_2(r) \D r^2+S^2(r) \D \Sigma_{k,n-2}^2\,, \qquad 
\phi=\phi(r) \,, 
\end{align}
where 
$\D \Sigma_{k,n-2}^2$ denotes the maximally symmetric space
with ${}^{(\gamma)}R_{ijkl}=2k\gamma_{i[k}\gamma_{l]j}$ with $k=0, \pm 1$.\footnote{
As far as the satisfaction of Einstein's equations are concerned, it gives no harm to suppose 
$\D \Sigma_{k,n-2}^2=\gamma_{ij}(z)\D z^i \D z^j$ to be the metric of Einstein space 
with ${}^{(\gamma)}R_{ij}=(n-3)k \gamma_{ij}$. However, we shall not attempt to do this here and confine to the maximally symmetric space for the sake of clarity. } 
A complete classification of the solutions of this form in Einstein-phantom-scalar system has been recently done in our accompanying paper \cite{paperI}, leading to a richer variety of solutions compared to the non-phantom case. 
We found that solutions fall into four family: the Fisher solution \cite{Fisher:1948yn}, the Ellis-Gibbons solution \cite{Ellis1973}, the Ellis-Bronnikov solution \cite{Ellis1973,Bronnikov1973} and the Erices-Mart\'inez solution \cite{Erices:2015xua}. Since the second and forth solutions are unified into a family of Gibbons solution \cite{Gibbons:2003yj,Gibbons:2017jzk}, we shall therefore consider  (i) the Fisher solution, (ii) the Gibbons solution and (iii) the Ellis-Bronnikov solution, 
despite that the general Gibbons solution falls out of the class (\ref{pseudospher}).  
For the physical discussion, we will limit ourselves to the asymptotically flat and spherically symmetric case. 

In appendix \ref{sec:curv}, we summarize the results of the curvature tensors for the 
metric (\ref{pseudospher}). 
On top of the conventional scalar quantities constructed out 
of the curvature tensors, the curvature tensors in a frame that is parallelly propagated
along the radial null geodesics are of crucial importance for the present analysis. 
We shall show that many solutions suffer from this kind of the p.p curvature singularities.

\subsubsection{Fisher solution}

First class of metric we consider is the $n(\ge 4)$-dimensional Fisher-class solution \cite{Fisher:1948yn} 
\begin{subequations}
\label{FJNW}
\begin{align}
\D s^2=&-f(r)^{\alpha}\D {t}^2+ f(r)^{-(\alpha+n-4)/(n-3)}\biggl(\D r^2+r^2f(r)\D \Sigma_{k,n-2}^2 \biggl),\\
\varphi=&\pm\frac{\sqrt{\epsilon(1-\alpha^2)}}{2a_{c}}\ln f(r), \qquad f(r)\equiv k-\frac{M}{r^{n-3}},
\end{align}
\end{subequations}
where $\alpha$ and $M$ are constants. 
This solution reduces to the Fisher solution in the spherically symmetric case in four dimensions ($n=4$ and $k=1$)~\cite{Fisher:1948yn}.
Its higher-dimensional generalization with $k=1$ was carried out by Xanthopoulos and Zannias~\cite{JNWhigher}. The parameter $\alpha $ takes values in $\alpha^2 \le 1 (\ge 1)$ for $\epsilon =+1$ ($\epsilon=-1$).

For $\alpha^2=1$, this Fisher-class solution reduces to the Schwarzschild-Tangherlini vacuum solution with general $k$. One may add a constant in the above expression of $\varphi$, but we have set it to zero for simplicity. As shown in our previous paper \cite{paperI}, the spherical Fisher solution with $\alpha^2\ne 1$ is necessarily singular, in that it inescapably admits a scalar curvature singularity or p.p curvature singularity. 
This can be deduced from the $R_{\mu\nu}k^\mu k^\nu$ component (\ref{Rkk}) of the Ricci tensor.

\subsubsection{Gibbons solution}

The next class of metrics we consider is the $n(\ge 4)$-dimensional Gibbons solution~\cite{Gibbons:2003yj}. 
This is a solution for the phantom case  only ($\epsilon=-1$) and is given by 
\begin{align}
\D s^2=-e^{-H}\D t^2+e^{H/(n-3)}h_{IJ}\D y^I\D y^J,\qquad \varphi=\pm \frac{H}{2a_c},
\label{Gibbons}
\end{align}
where $h_{IJ}$ is an arbitrary $(n-1)$-dimensional Ricci-flat Riemannian metric ${}^{(n-1)}{R}_{IJ}=0$ and $H$ is a harmonic function $\Delta_h H=0$ on $M^{n-1}$. We have neglected a constant which may be added to  the above expression of $\varphi$. 

A distinguished feature of the Gibbons solution (\ref{Gibbons}) is that the system is completely linearized since $H$ obeys the harmonic equation. 
This  is reminiscent of the Majumdar-Papapetrou solution \cite{Majumdar:1947eu,Papaetrou:1947ib}
in the Einstein-Maxwell system. The linearization of the field equations in the present sytem  
does not occur for the non-phantom case. 
To gain a better perspective into this, let us go back to the sigma model equations (\ref{EOMNLSM1}). 
For the phantom ($ \epsilon=-1$) case, the target space is $\D s_T^2=\D \sigma^2 -\D \varphi^2$. 
Setting $\varphi=\pm \sigma$, we have a null direction for the target space 
$\D s_T^2=G_{ab}\D \phi^a \D \phi^b=0$, for which 
equations of motion (\ref{EOMNLSM1}) imply that the base space is Ricci flat ${}^{(n-1)}R_{IJ}=0$
and $\Delta_h \sigma =0$, giving rise to the Gibbons solution (\ref{Gibbons}). 
For the non-phantom case, the target space is Riemannian $\D s_T^2=\D \sigma^2 +\D \varphi^2$.
This prohibits the null direction of the target space, for which the nonlinearity persists. 
It turns out that the possibility of the complete linearization is intrinsic to the phantom case.

If we limit ourselves to the flat base space 
and  allow only for the spherical harmonics, the Gibbons solution (\ref{Gibbons}) falls 
into the class of metrics under study (\ref{pseudospher}). 
Under these conditions, the solution (\ref{Gibbons}) reduces to 
\begin{align}
\label{EllisGibbons}
\D s^2=&-e^{-H(r)}\D t^2+e^{H(r)/(n-3)}\bigl(\D r^2+r^2\D \Omega_{n-2}^2\bigl),\qquad 
\varphi=\pm \frac{H(r)}{2a_c}\,, 
\end{align}
where $H(r)=M/r^{n-3}$ is the monopole harmonic function with $M$ being a constant. 
$\D \Omega_{n-2}^2$ is the standard metric of the unit $(n-2)$-sphere. 
The metric in~(\ref{EllisGibbons}) in $n=4$ has been also investigated in~\cite{Yilmaz} for the purpose to propose an alternative theory for gravity.

As demonstrated in a companion paper, we have shown that the Ellis-Gibbons solution (\ref{EllisGibbons})
is always singular at $r=0$, regardless of the sign on $M$ and the spacetime dimensionality. 
The divergence of the scalar curvature only occurs for $M<0$. For $M>0$,  $r=0$ becomes a p.p curvature singularity. Moreover, the $r=0$ surface for $M>0$ is achievable within a finite affine parameter 
for radial null geodesics, even though the areal radius is diverging there. 
These properties rule out the possibility that the Eills-Gibbons solution (\ref{EllisGibbons}) describes the regular wormhole spacetime, contrary to the claim in the literature \cite{Boonserm:2018orb}.

\subsubsection{Ellis-Bronnikov solution}

The third class of solutions is the $n(\ge 4)$-dimensional Ellis-Bronnikov solution~\cite{Ellis1973,Bronnikov1973}. 
This solution exists only in the phantom case ($\epsilon=-1$) with $k=1$ and is given by
\begin{subequations}
\begin{align}
\label{Ellis-Bronnikov}
\D s^2&=-e^{-2\beta U(r)}\D t^2+e^{2\beta U(r)/(n-3)}V(r)^{1/(n-3)}\biggl(\frac{\D r^2}{V(r)} +r^2
\D \Omega_{n-2}^2 \biggl),\\
\varphi&=\pm\frac{\sqrt{1+\beta^2}}{a_{c}}U(r)\,,
\end{align}
\end{subequations}
with
\begin{align}
\label{}
U(r)\equiv  \arctan \left(\frac{M}{2r^{n-3}}\right)\,, \qquad V(r)\equiv  1+\frac{M^2}{4r^{2(n-3)}} \,,
\end{align}
where $\beta$ and $M$ are parameters.
Since the metric is invariant under $M\to -M$ and $\beta \to -\beta$, one can restrict to the 
$M>0$ case.

As explained in our previous paper \cite{paperI}, the Ellis-Bronnikov solution in $n\ge 4$ dimensions deserves a 
traversable wormhole which bridges the two asymptotically flat regions. The solution does not allow any spacetime points where the scalar curvature invariants and 
the Riemann tensor components in a parallelly propagated frame 
become unboundedly large. 
The $r=0$ surface is therefore a mere coordinate singularity. The extension through 
$r=0$ is best achieved by the replacement $U(r)\to\pi/2-\arctan(2r^{n-3}/M)$ and $x=r^{n-3}$
(use $\arctan(x)+\arctan(1/x)=\pi/2$ for $x>0$). The resulting metric is smooth at $x=0$ and can be extended into the $x<0$ region. This yields a maximal extension of the spacetime with two asymptotically flat regions, i.e., the spacetime describes a regular wormhole. 
%See \cite{Clement:1983fe} for wormholes coupled to multiple scalar fields and \cite{Bronnikov:1997gj} for ring-like wormholes. 

%
%We now turn our attention to the $E\ne 0$ case, which will be divided into two cases 
%according to $a_c^2+\epsilon a^2=0$ or not. We shall refer to the case 
%with $a=a_c$ with $\epsilon=-1$ as critical coupling, otherwise as non-critical coupling. 
%
%
%

\section{Noncritical case}
\label{sec:noncritical}

We are now in a position to formulate the solution-generating transformations in the system (\ref{action-dil2}) based on the symmetry of the reduced action (\ref{harmonicmap}) with a non-critical case 
$a^2+\epsilon a_c^2\ne 0$.
For $\epsilon=1$, transformation rules in the Einstein-Maxwell theory have been derived for  $n=4$ in~\cite{jrw1969}, for $a=0$ arbitrary $n(\ge 4)$ in~\cite{Maeda:2019tqs}. 
The Einstein-Maxwell-dilaton gravity for $n(\ge 4)$ has been argued in~\cite{Yazadjiev:2005hr}.
See \cite{Kinnersley:1977pg} for the discussion in more general nonstatic context for $a=0$ 
and \cite{Galtsov:1995mb} for $a\ne 0$.
See also~\cite{Galtsov:1998mhf} for the $p$-dimensional reduction of the system with a $p$-form field and 
\cite{Huang:2019arj} for a different kind of dilaton coupling.

\subsection{Solution-generating transformations}

In the case of the non-critical coupling satisfying $a^2+\epsilon a_{c}^2 \ne 0$, we can adopt new variables $\phi_{1-3}$ by 
\begin{align}
\label{phivar}
\sigma=&\frac{a\phi_1+{\epsilon}a_{c}\phi_2}
{a^2+\epsilon a_{c}^2}\,, \qquad 
\varphi=\frac{a_{c}\phi_1-a\phi_2}{a^2+\epsilon a_{c}^2}\,, \qquad 
E=C\phi_3 \,, 
\end{align}
where 
\begin{align}
\label{}
C\equiv \frac{1}{\sqrt{a^2+\epsilon a_c^2}}\,.
\end{align}
In terms of these variables, the target space metric (\ref{targetmet}) is transformed into 
a more recognizable form
\begin{align}
\label{target:noncritical}
\D s_T^2=2C^2 \left(\D\phi_1^2+\epsilon \D\phi_2^2-e^{2\phi_2 }\D \phi_3^2\right) \,.
\end{align}
For $\epsilon=1$, this target space is $\mathbb R\times {\rm SL}(2,\mathbb R)/{\rm SO}(1,1)$. 
For $\epsilon=-1$, in contrast, the sign of $C^2$ may be indefinite but the coset structure is 
$\mathbb R\times {\rm SL}(2,\mathbb R)/{\rm SO}(2)$ insensitive to the sign of $C^2$. 
A standard scheme for finding the finite transformations for the symmetric space is to construct the 
coset representative of the space (see e.g., \cite{Breitenlohner:1987dg} for the conceptual foundations of the coset construction). However, we do not bank on this strategy using the coset representative, since the case-by-case analyses for $\epsilon =\pm 1$ are inevitable.  
Instead, we will go on in a very pedestrian way by the direct integration of Killing's equations for the original metric (\ref{targetmet}). Upon exponentiation of the obtained Killing vectors, we get the corresponding finite transformations.

An elementary practice gives rise to all the four Killing vectors $\xi_{(A)}~(A=1,2,3,4)$ of the target space (\ref{targetmet}) as (see~\cite{Gibbons:1996pd} for $n=4$)
\begin{align}
\xi_{(1)}=&\,\partial_\sigma-a_c E \partial_E \,, \qquad 
\xi_{(2)}=\partial_\varphi+a E \partial_E \,, \qquad 
\xi_{(3)}=\partial_E \,,\notag  \\
\xi_{(4)} =&\,\epsilon E \partial_\sigma -\frac{a}{a_c } E \partial_\varphi -
\frac{E^2(a^2+\epsilon a_c^2)+\epsilon e^{2a \varphi-2a_c \sigma}}{2a_c}\partial_E \,. 
\label{KVsnoncrit}
\end{align}
These Killing vectors constitute the 
 $\mathfrak{gl}(2,\mathbb R)=\mathbb R\oplus \mathfrak{sl}(2,\mathbb R)$ algebra
\begin{align}
\label{}
[\xi_{(1)}, \xi_\pm ]=a_c \xi_\mp \,, \qquad 
[\xi_{(2)},  \xi_\pm]=-a \xi_\mp  \,, \qquad 
[\xi_{(3)}, \xi_{(4)}]=\epsilon \xi_{(1)}-\frac{a}{a_c}\xi_{(2)} \,,
\end{align}
where $\xi_\pm\equiv \xi_{(3)}\pm \xi_{(4)}$. 
One finds that $a\xi_{(1)}+a_c \xi_{(2)}$ commutes with other generators and embodies the $\mathbb R$ generator. 
The finite transformations $\{\sigma,\varphi,E\}\to \{\sigma',\varphi',E'\}$ corresponding to the Killing vectors $\xi_{(1)}$--$\xi_{(4)}$ are respectively given by
\begin{subequations}
\begin{align}
{\rm [I]}~~& \sigma'=\sigma+c_1, \qquad \varphi'=\varphi \,, \qquad E'=e^{-a_c c_1} E,\label{non-criticalI}\\
{\rm [II]}~~& \sigma'=\sigma,\qquad  \varphi'=\varphi+c_2 \,, \qquad E'=e^{a c_2} E\,, \label{non-criticalII}\\
{\rm [III]}~~& \sigma'=\sigma,\qquad  \varphi'=\varphi,\qquad  E'=E+c_3 \,, \label{non-criticalIII}\\
{\rm [IV]}~~& \sigma'=\sigma+\frac{a_c \epsilon}{a^2+a_c^2 \epsilon}\ln\Xi'\,,\qquad \varphi'=\varphi-\frac{a}{a^2+a_c^2 \epsilon}\ln\Xi' \,,\notag \\
& E'=\frac{1}{\Xi'}\left[E+\frac{c_4}{2a_c}
\left\{{(a^2+{\epsilon}a_c^2)E^2}-\epsilon e^{2a\varphi-2a_c \sigma}\right\}\right]
\,,\label{non-criticalIV}
\end{align}
\end{subequations}
where $c_1$--$c_4$ are constants and 
\begin{align}
\label{}
\Xi'\equiv  \left(1+\frac{c_4 E(a^2+\epsilon a_c^2)}{2a_c}\right)^2-\frac{\epsilon c_4^2(a^2+\epsilon a_c^2)
e^{2a\varphi-2a_c \sigma}}{4a_c^2}\,.
\end{align}
The transformations I---III are pure gauge transformations:
the constant shift of the norm of the timelike Killing vector [I] and that of the dilaton [II] together with a constant scaling of electrostatic potential are compensated by the rescaling of the spacetime coordinates ($t, y^I$); 
[III] is nothing but the choice of the origin for the electrostatic potential. 
 The transformation [IV] is only nontrivial and allows one to obtain a charged solution from 
a neutral solution, corresponding to the Harrison transformation \cite{Harrison}
in Einstein-Maxwell theory in four dimensions. 
It is worth commenting that the above transformation [IV] does not generate any scalar charge
from the $\varphi=0$ solution in the $a=0$ case.

Now let us apply the constructive procedure I--IV given by (\ref{non-criticalI})--(\ref{non-criticalIV}) to generate new exact solutions.
We assume that the seed solution $\phi^a_0=\{\sigma_0, \varphi_0, E_0\}$ is electrically neutral 
$(E_0=0)$ and satisfies the boundary condition $(\sigma_0, \varphi_0) \to 0$ as $r\to \infty$.
By applying the transformations (\ref{non-criticalI})--(\ref{non-criticalIV})  successively and choosing the constants suitably, one finds the new solution $\{\sigma, \varphi, E\}$ preserving the boundary condition as
\begin{align}
\label{chargetr:noncrit}
\sigma& =\sigma_0+\frac{\epsilon a_c}{a^2+\epsilon a_c^2}\ln \Xi \,, \qquad 
\varphi=\varphi_0-\frac{a}{a^2+\epsilon a_c^2}\ln \Xi \,, \qquad
E =\frac{q (e^{2a\varphi_0-2a_c \sigma_0}-1) }{\{1-\epsilon q^2(a^2+\epsilon a_c^2)\}\Xi}\,,
\end{align}
where
\begin{align}
\label{}
\Xi \equiv  \frac{1-\epsilon q^2 e^{2a\varphi_0-2a_c \sigma_0} (a^2+\epsilon a_c^2)}{1-\epsilon q^2(a^2+\epsilon a_c^2) }.
\end{align}
Here $q$ is a new parameter corresponding to the electric charge.

Using the formulae (\ref{chargetr:noncrit}), we will next derive explicit forms of static charged solutions. 
For the spherical case ($k=1$), the boundary condition imposed above amounts to the asymptotic flatness. 
One can similarly impose the boundary conditions on the the topological version of the Fisher solution, 
but we shall not attempt to do this here and we specialize only to the asymptotically flat case. 
In the presence of a dilaton coupling ($a\ne 0$), it is also possible to generate non-asymptotically flat 
solutions by the transformations I--IV~\cite{Chan:1995fr,kp1997,Yazadjiev:2005du,Yazadjiev:2005hr}. However, we also preclude this possibility from our analysis.

\subsection{Charged Fisher solution}

The functions $\sigma=\sigma_0$ and $g_{IJ}$ in the metric (\ref{metric}) and the scalar field $\varphi=\varphi_0$ for the spherical Fisher seed solution (\ref{FJNW}) are given by  
\begin{align}
\label{JNWseed}
&\varphi_0=\pm \frac{\sqrt{\epsilon(1-\alpha^2)}}{2a_c}\ln f \,, \qquad 
\sigma_0=-\frac{\alpha}{2a_c}\ln f\,, \notag \\
&g_{IJ}\D y^I \D y^J =f^{-(n-4)/(n-3)}\left(\D r^2+r^2 f\D \Omega_{n-2}^2 \right) \,,
\end{align}
where $f=1-M/r^{n-3}$. 
By the transformation (\ref{chargetr:noncrit}), we obtain the following asymptotically flat 
charged solution
\begin{subequations}
\begin{align}
\label{}
\D s^2 =&-f^\alpha {\Xi_{\rm F}}^{\alpha_1} \D t^2+
f^{-(\alpha+n-4)/(n-3)} {\Xi_{\rm F}}^{-\alpha_1/(n-3)}\left(\D r^2+r^2 f\D \Omega_{n-2}^2\right) \,, \\
\varphi=& \pm\frac{\sqrt{\epsilon(1-\alpha^2)}}{2a_c}\ln f -\frac{a}{a^2+\epsilon a_c^2}\ln \Xi _{\rm F}\,, 
\qquad 
E=\frac{q}{1-{\epsilon}q^2(a^2+\epsilon a_c^2)}\frac{f^{\alpha_2}-1}{\Xi_{\rm F}}\,,
\end{align}
\end{subequations}
where 
\begin{align}
\label{}
\Xi_{\rm F}\equiv  \frac{1-\epsilon q^2(a^2+\epsilon a_c^2)f^{\alpha_2}}{1-\epsilon q^2(a^2+\epsilon a_c^2)}\,, \qquad 
\alpha_1\equiv -\frac{2\epsilon a_c^2}{a^2+\epsilon a_c^2}\,, \qquad 
\alpha_2\equiv \alpha\pm \sqrt{\epsilon(1-\alpha^2)}\frac{a}{a_c} \,.
\end{align}
We examine the physical properties of this solution in the following.

\subsubsection{$\alpha^2=1$}

Let us first consider the $\alpha^2=1$ case. 
Note first that the $\alpha=-1$ case is obtainable by $M\to -M$ with $r^{n-3}\to r^{n-3}+M$, so that we consider the $\alpha=1$ case only. This solution should recover the one in~\cite{Gibbons:1987ps}  for $\epsilon=1$ and the one  in~\cite{Gao:2006iw} for $\epsilon=-1$. 
To see this, let us redefine $r^{n-3}\to r^{n-3} -Mq^2 \epsilon(a^2+a_c^2\epsilon)/[1-q^2 \epsilon(a^2+a_c^2 \epsilon)]$ and set $M=r_+^{n-3}-r_-^{n-3}$, 
$q=\sqrt{r_-^{n-3}/[r_+^{n-3}\epsilon(a^2+a_c^2\epsilon)]}$. Then the solution reduces to 
\begin{subequations}
\label{FisherBH}
\begin{align}
 \D s^2_{\alpha=1}=& -f_+(r)f_-(r)^{-(a^2-a_c^2\epsilon)/(a^2+a_c^2\epsilon)}\D t^2
 +f_-(r)^{2a^2/[(n-3)(a^2+a_c^2\epsilon)]}\left(
 \frac{\D r^2}{f_+(r)f_-(r)}+r^2 \D \Omega_{n-2}^2 
 \right) \,,  \\
 \varphi=& \frac{a}{a^2+a^2_c\epsilon}\ln f_-(r)\,, \qquad 
 E=-\frac{r_+^{n-3}}{r^{n-3}}\sqrt{\frac{r_-^{n-3}}{r_+^{n-3}\epsilon(a^2+a_c^2\epsilon)}} \,, 
\end{align}
\end{subequations}
where $f_\pm(r) \equiv 1-r_\pm^{n-3} /r^{n-3}$. This expression matches well with the ones in \cite{Gibbons:1987ps,Gao:2006iw}, as we desired to show. 
In the above form of the metric, we do not need to demand the positivity of $r_\pm^{n-3}$ 
nor the relation $r_+>r_->0$. The only restriction  comes from the reality of $q$, viz, 
\begin{align}
\label{Fishercond}
\frac{r_-^{n-3}}{r_+^{n-3}\epsilon(a^2+a_c^2\epsilon)}>0 \,. 
\end{align}
The ADM mass reads
\begin{align}
\label{}
\ma M=\frac{(n-2)\Omega_{n-2}}{2\kappa_n} \left(r_+^{n-3}-\frac{a^2-a_c^2\epsilon}
{a^2+a_c^2\epsilon}r_-^{n-3}\right)\,. 
\end{align}
The electric charge is defined by 
\begin{align}
\label{}
\ma Q \equiv  -\frac{1}{\kappa_n}\int e^{-2a \phi} F_{\mu\nu}\D S^{\mu\nu} \,, 
\end{align}
leading to 
\begin{align}
\label{}
\ma Q=\frac{2(n-3)\Omega_{n-2}r_+^{n-3}}{\kappa_n}\sqrt{\frac{r_-^{n-3}}{r_+^{n-3}\epsilon(a^2+a_c^2\epsilon)}}
\,.
\end{align}

Let us first look into the regularity of the $r=r_-$ surface. 
The leading-order term of the Ricci scalar reads 
$R\sim  f_-^{-\delta-2a^2/[(n-3)(a^2+a_c^2\epsilon)]}$ around $r=r_-$, 
where $\delta=1$ for $r_+\ne r_-$ and $\delta =0$ for $r_+=r_-$. 
Other curvature invariants display the analogous behavior. 
Thus the curvature invariants remain finite at $r=r_-$ only for 
the phantom case with $a_c\sqrt{(n-3)/(n-1)}\delta \le a<a_c$. 
We stress that one cannot immediately conclude that the $r=r_-$
is a regular surface in this parameter range. 
To see this, we evaluate the p.p frame Ricci tensor  (c.f (\ref{Rkk}))
\begin{align}
\label{}
R_{\mu\nu}k^\mu k^\nu = \frac{a^2a_c^2(n-3)(n-2)r_-^{2(n-3)}\epsilon}{(a^2+a^2_c\epsilon)^2 r^{2(n-2)}}
f_-^{-2+\frac{2a^2(n-4)}{(n-3)(a^2+a_c^2\epsilon)}} \,, 
\end{align}
where 
$k^\mu=f_+^{-1}f_-^{(a^2-a_c^2\epsilon)/(a^2+a_c^2\epsilon)}(\partial_t )^\mu+f_-^{(n-4)a^2/[(n-3)(a^2+a_c^2\epsilon)]}(\partial_r)^\mu$
is an affine parameterized radial null geodesic tangent $k^\nu\nabla_\nu k^\mu=0$. In the phantom case, 
$R_{\mu\nu}k^\mu k^\nu $ tends to be finite as $r\to r_-$ only for 
$a_c<a\le \sqrt{n-3}a_c$. This covers the complementary range in which 
the curvature invariants are finite. 
We therefore conclude that $r=r_-$ is singular in any case. 

Repeating the same calculation for $r_-^{n-3}<0$, the $r=0$ surface is always singular. The only regular configuration admitting the horizon is therefore the $r_+^{n-3}>0$ case with 
$r_+^{n-3}>{\rm max}(0, r_-^{n-3})$. The $r=r_+(>0)$ surface deserves an event horizon of a black hole, since the regularity and the finiteness of the affine parameter for the radial null geodesics are satisfied. 
In this case, the surface gravity $\kappa$, the area $A_{\rm H}$, and the electrostatic potential 
$\Phi_{\rm H}$ of the horizon are given by
\begin{subequations}
\begin{align}
\label{surfacegrav}
\kappa= &\frac{n-3}{2r_+} \left(1-\frac{r_-^{n-3}}{r_+^{n-3}}\right)^{[-a^2+(n-3)a_c^2\epsilon]
/[(n-3)(a^2+a_c^2\epsilon)]} \,,  \\
A_{\rm H}=& \Omega_{n-2}r_+^{n-2}  \left(1-\frac{r_-^{n-3}}{r_+^{n-3}}\right)^{(n-2)a^2/[(n-3)
(a^2+a_c^2\epsilon)]}\,, \\
\Phi_{\rm H}=&\sqrt{\frac{r_-^{n-3}}{r_+^{n-3}\epsilon(a^2+a_c^2\epsilon)}}\,, 
\end{align}
\end{subequations}
satisfying the first law of black hole thermodynamics
\begin{align}
\label{1st}
\delta \ma M=\frac{\kappa}{\kappa_n }\delta A_{\rm H}+\Phi_{\rm H} \delta \ma Q \,.
\end{align}

In the degenerate limit $r_+=r_-$ with $\epsilon=-1$, the solution in $n=4$ reduces to (3.37) of \cite{Clement:2009ai}. 
The authors in \cite{Clement:2009ai} imposed the analyticity of the metric there and concluded that this solution with 
$\epsilon=-1$ and $a^2=(p-1)/(p+1)$ ($p\in \mathbb N$) yields 
an asymptotically flat black hole which is regular on and outside the event horizon. 
Albeit the finiteness of scalar curvature invariants for $a<a_c$, 
the $r=r_- $ surface is  the p.p curvature singularity as demonstrated above. 
It follows that the degenerate limit $r_+\to r_- (>0)$ of the solution (\ref{FisherBH}) fails 
to describe a black hole with a regular event horizon, in opposition to the results in \cite{Clement:2009ai}.

Let us next elaborate on the inner structure of the black hole $r_+^{n-3}>0$.
The analysis for the singularity can be 
divided into (i) $\epsilon=+1$ and (ii) $\epsilon=-1$. 
The $\epsilon=-1$ case is further categorized into 
(ii-a) $a<a_c$, (ii-b) $a_c <a\le \sqrt{(2n-5)/(n-3)}a_c$ and 
(ii-c) $\sqrt{(2n-5)/(n-3)}a_c<a$. For case (i), (\ref{Fishercond})  gives 
$r_+>r_->0$ and our criterion (\ref{tortoise}) implies that 
$r=r_-$ is spacelike. 
In case (ii-a), we have $r_+>r_->0 $ from (\ref{Fishercond})  and 
we find that $r=r_-$ is null. 
In case (ii-b,c), we have $r_-^{n-3}<0$ and that 
$r=0$ is null for  (ii-b) and spacelike for (ii-c). 
Using (\ref{affine}), we can verify that the affine parameter along the radial null geodesics 
to the singularity (either $r=0$ or $r=r_-$) is finite.

\begin{figure}[t]
\begin{center}
\includegraphics[width=14cm]{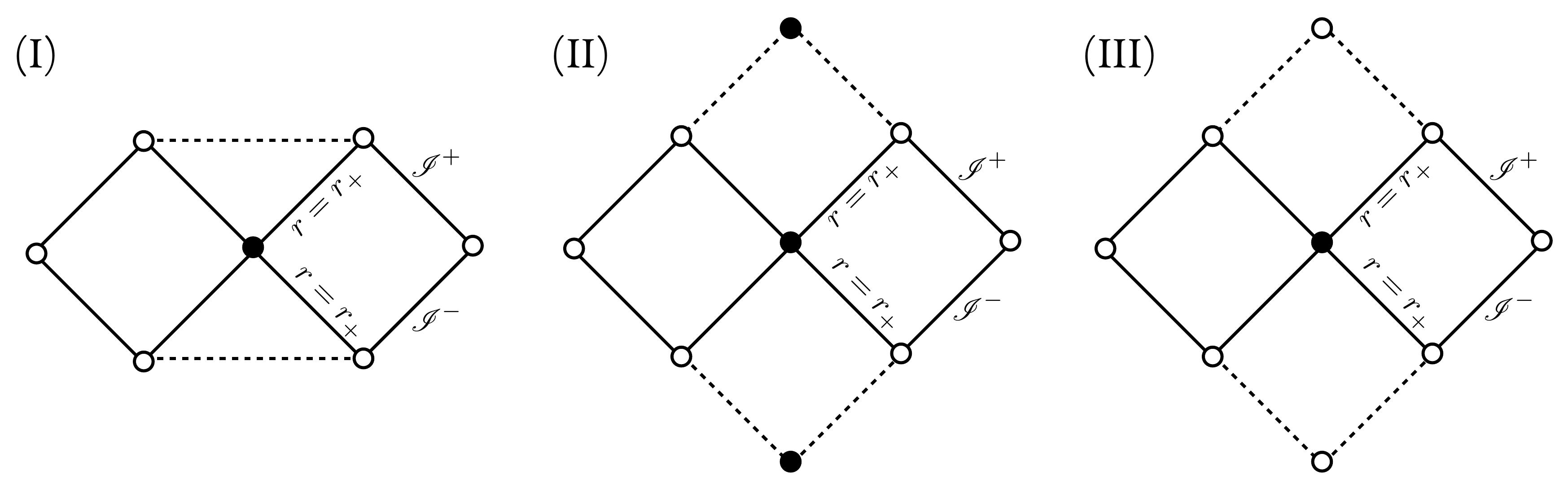}
\caption{Possible conformal diagrams for the charged Fisher solution 
with a horizon for $\alpha^2=1$ (critical and noncritical cases). 
Dashed lines correspond to the (scalar or p.p) curvature singularities. 
White and black circles stand  for spatial/timelike infinities, 
and for the bifurcation surface, respectively.}
\label{fig:PDFisher}
\end{center}
\end{figure}

In the interior of the black hole, $r$ plays the role of the time coordinate. 
Let us finally investigate the proper time along the timelike geodesics 
toward the singularity. This can be evaluated as 
\begin{align}
\label{}
\tau = \int ^r \frac{f_-(r)^p}{|f_+(r)|^{1/2}}\D r \,, \qquad 
p=-\frac{(n-5)a^2+(n-3)a_c^2\epsilon}{2(n-3)(a^2+a_c^2\epsilon)} \,. 
\end{align}
In the integration diverges as $r\to 0$ or $r\to r_-$, an infinite proper time elapses to reach the singularity. 
The results are summarized in table \ref{table:FisherBH}. Combining the results obtained in this section, we arrive at the possible Penrose diagrams, which are shown in (\ref{fig:PDFisher}). 

%-------------- TABLE ---------------%
\begin{table}[t]
\begin{center}
\caption{
The properties of the singularity for the 
parameter region under which the charged Fisher solution with $\alpha^2= 1$ 
(\ref{FisherBH}) admits a black hole horizon at $r=r_+(>0)$.}
\vspace{0.2cm}
\scalebox{0.75}{
%\small 
\begin{tabular}{c|c|c|c|c|c|c}
\hline 
kinetic term &$\epsilon=1$ & \multicolumn{5}{|c}{$\epsilon=-1$}   \\ \hline\hline
dilaton coupling & arbitrary & $a<\sqrt{\frac{n-3}{n-1}}a_c$ &  $\sqrt{\frac{n-3}{n-1}}a_c\le a<a_c$ & $a_c<a\le \sqrt{\frac{n-2}{n-3}}a_c $ & $\sqrt{\frac{n-2}{n-3}}a_c<a\le\sqrt{\frac{2n-5}{n-3}}a_c$ &
$\sqrt{\frac{2n-5}{n-3}}a_c<a$ \\ \hline 
singularity & $r=r_-(>0)$ & \multicolumn{2}{|c}{$r=r_-(>0)$} &  \multicolumn{3}{|c}{$r=0$~($r_-^{n-3}<0$)}   \\ \hline
signature &spacelike & null & null & null & null & spacelike\\ \hline 
null affine distance & finite &  finite&  finite&  finite & finite  & finite \\ \hline
proper time &finite & finite & infinite & infinite & finite & finite \\ \hline
\begin{tabular}{c}
Penrose diagram\\
 in figure \ref{fig:PDFisher}
 \end{tabular} & (I) & (II) &(III) & (III) &(II) & (I)   \\ \hline 
\end{tabular} 
}
\label{table:FisherBH} 
\end{center}
\end{table} 
%------------------------------------%

\subsubsection{$\alpha^2\ne 1$}

Let us next analyze the $\alpha^2\ne 1$ case, in which the plus and minus branches are 
exchanged under $\alpha \to -\alpha $ with $M\to -M$. This permits us to focus on the plus branch. 
The Ricci scalar and the Ricci tensor component have the following structure
\begin{subequations}
\begin{align}
\label{}
R&=\frac{f^{\frac{2-n+\alpha}{n-3}}\Xi^{\frac{6-2n+\alpha_1}{n-3}}}{r^{2(n-2)}[1-q^2(a_c^2+\epsilon a^2)]^2}
(A_1+A_2 q^2 f^{\alpha_2}+A_3q^4 f^{2\alpha_2})\,, \\
R_{\mu\nu}k^\mu k^\nu&=\frac{f^{-2-\frac{(n-4)(\alpha-1)}{n-3}}\Xi^{-2-\frac{(n-4)\alpha_1}{n-3}}}{r^{2(n-2)}[1-q^2(a_c^2+\epsilon a^2)]^2}
(B_1+B_2 q^2 f^{\alpha_2}+B_3q^4 f^{2\alpha_2})\,,
\end{align}
\end{subequations}
where $A_{1-3}$ and $B_{1-3}$ are  $q$-independent constants which are
generically nonvanishing for $\alpha^2\ne 1$. 
We see that either of these quantities diverge at each surface $f=0$, $r=0$ (and $\Xi=0$ if any). 
It follows that the solution with $\alpha^2\ne 1$ describes a singular spacetime.

\subsection{Charged Gibbons solution}

The functions $\sigma=\sigma_0$ and $g_{IJ}$ in the metric (\ref{metric}) and the scalar field $\varphi=\varphi_0$ for this seed solution (\ref{Gibbons}) are given by  
\begin{align}
\label{seed:Gibbons}
&\varphi_0=\pm \frac{H}{2a_c}\,, \qquad 
\sigma_0=\frac{H}{2a_c}\,, \qquad g_{IJ} =h_{IJ} \,,
\end{align}
where $h_{IJ}$ is the Ricci-flat metric and $H$ is a harmonic function $\Delta_h H=0$ thereof. 
The transformation (\ref{chargetr:noncrit}) with $\epsilon=-1$ gives rise to the charged Gibbons solution
\begin{subequations}
\begin{align}
\label{}
\D s^2 &=-e^{-H} {\Xi_{\rm G}}^{2a_c^2/(a^2-a_c^2)} \D t^2+  e^{H/(n-3)} {\Xi_{\rm G}}^{-2a_c^2/[(n-3)(a^2-a_c^2)]}
 h_{IJ}\D y^I \D y^J \,, \\
\varphi&=\pm\frac{H}{2a_c}-\frac{a}{a^2-a_c^2}\ln \Xi_{\rm G}\,, \qquad 
E=\frac{q[e^{(\pm a-a_c)H/a_c}-1]}{[1+q^2(a^2-a_c^2)]\Xi_{\rm G}}\,, 
\end{align}
\end{subequations}
where
\begin{align}
\label{}
\Xi_{\rm G}\equiv  \frac{1+q^2(a^2-a_c^2)e^{(\pm a-a_c)H/a_c}}{1+q^2(a^2-a_c^2)} \,. 
\end{align}
This solution with $a=0$ has been derived in~\cite{Maeda:2016ddh}. 
In what follows, we look into the properties of the solution by 
focusing on the spherically symmetric case, i.e., $h_{IJ}\D y^I\D y^J=\D r^2+r^2 \D \Omega_{n-2}^2$
and $H=M/r^{n-3}$. 
%The plus and minus branch solutions are then switched around by a simultaneous sign flip $a\to -a$ with $\varphi\to -\varphi$. Here we use this symmetry to fix $a>0$ and consider each branch. 
The Ricci scalar and the p.p frame of the Ricci tensor are then given by 
\begin{align}
\label{}
R=&-\frac{(n-3)M^2}{r^{2(n-2)}}e^{-H/(n-3)}\Xi^{-2+\frac{2a_c^2}{(n-3)(a^2-a_c^2)}}
\Biggl(\frac{n-2}4\notag \\
&+\frac{n-6}{2}(a\mp a_c)^2q^2 e^{(\pm a-a_c)H/a_c}+\frac{n-2}4q^4
(a\mp a_c)^4e^{2(\pm a-a_c)H/a_c}\Biggr)\,,
\end{align}
and 
\begin{align}
\label{}
R_{\mu\nu}k^\mu k^\nu=& -\frac{(n-2)(n-3)M^2}{4[1+q^2(a^2-a_c^2)]^2 r^{2(n-2)}}
e^{-\frac{n-2}{n-3}H}\Xi^{-2-\frac{2a_c^2(n-4)}{(n-3)(a^2-a_c^2)}}
%\notag \\ & \times 
\left(e^H-(a\mp a_c)^2q^2 e^{\pm a H/a_c} \right)^2 \,. 
\end{align}

Keeping $a>0$ in mind,\footnote{
Since the plus and minus branches are interchanged by simultaneous sign flip of 
$\varphi$ and $a$, one can restrict to either of the branches with $a\in \mathbb R$.
In the present case, it is more convenient to fix $a>0$ and consider both branches.} 
let us first examine the plus branch of the solution. 
We have four cases to consider: (i) $0<a<a_c$ with $M>0$, 
(ii) $a_c<a$ with $M>0$, (iii) $0<a<a_c$ with $M<0$ and 
(iv) $a_c<a$ with $M<0$. In case (i),  there appears a surface 
$r=r_+=[(a_c-a)M/(a_c \log[q^2(a_c^2-a^2)])]^{1/(n-3)}>0$ where 
$\Xi_{\rm G}(r_+)=0$. This surface is singular because of 
$R\propto \Xi^{-[2(n-2)(a_c^2-a^2)+2a^2]/[(a_c^2-a^2)(n-3)]}$. 
%From (\ref{affine}), the affine parameter for radial null geodesics is infinite
%for $a_c/\sqrt{n-3}\le a<a_c$.  Eq. (\ref{tortoise}) implies that this surface is timelike.
In case (ii), all curvature invariants seem to be finite at $r=0$, but the p.p frame component of the Ricci tensor $R_{\mu\nu}k^\mu k^\nu \propto -e^{(a-a_c)(n-4)M/[(a+a_c)(n-3)r^{n-3}]}/r^{2(n-2)}$ 
tends to blow up as $r\to 0$, implying the p.p curvature singularity. 
In case (iii), the Ricci scalar diverges 
$R \propto -e^{(a-a_c)M/[(a+a_c)(n-3)r^{n-3}]}/r^{2(n-2)}$ as $r\to 0$, while 
in case (iv)  $R\propto -e^{-M/[(n-3)r^{n-3}]}/r^{2(n-2)}$ as $r\to 0$. 
It thus follows that the plus branch solution is always singular.

Let us next consider the minus branch of the solution (with $a>0$).
We consider cases (i)--(iv) as the plus branch solution. 
In case (i), the $\Xi_{\rm G}=0$ surface appear in the $r>0$ region, which 
turns out to be singular. 
In case (iv), 
the Ricci scalar diverges 
$R\propto -e^{(a+a_c)M/[(a_c-a)(n-3)r^{n-3}]}/r^{2(n-2)} \to \infty$,
and remains finite in other cases (ii) and (iii). 
The Ricci tensor in a parallelly propagated frame reads
$R_{\mu\nu}k^\mu k^\nu \propto -e^{(n-4)M/[(n-3)r^{n-3}]}/r^{2(n-2)} $ for case (ii) 
and $R_{\mu\nu}k^\mu k^\nu \propto -e^{(a+a_c)(n-4)M/[(a-a_c)(n-3)r^{n-3}]}/r^{2(n-2)}$
for case (iii) as $r\to 0$. To conclude, 
the minus branch solution is also singular in any parameter region.

\subsection{Charged Ellis-Bronnikov solution}
\label{sec:dilatonEB}

The functions $\sigma=\sigma_0$ and $g_{IJ}$ in the metric (\ref{metric}) and the scalar field $\varphi=\varphi_0$ for this seed solution (\ref{Ellis-Bronnikov}) are given by  
\begin{align}
\label{seed:Ellis}
\sigma_0 =\frac{\beta}{a_c}U \,, \qquad 
\varphi_0=\pm\frac{\sqrt{1+\beta^2}}{a_{c}}U \,, \qquad 
g_{IJ}\D y^I \D y^J=V^{1/ (n-3)}\left(\frac{\D r^2}{V}+r^2\D \Omega_{n-2}^2\right)\,,
\end{align} 
where $V=1+M^2/(4r^{2(n-3)})$ and $U=\arctan[M/(2r^{n-3})]$. 
By the transformation (\ref{chargetr:noncrit}) with $\epsilon=-1$, 
we derive the following charged Ellis-Bronnikov solution
\begin{subequations}
\begin{align}
\label{dilaton-EB}
\D s^2 &=-e^{-2\beta U} {\Xi_{\rm EB}}^{2a_c^2/(a^2-a_c^2)} \D t^2+  \left[e^{2\beta U}
{\Xi_{\rm EB}}^{-2a_c^2/(a^2-a_c^2)}V\right]^{1/ (n-3)}\left(\frac{\D r^2}{V}+r^2\D \Omega_{n-2}^2\right)\,, \\
\varphi&=\pm\frac{\sqrt{1+\beta^2}}{a_c} U-\frac{a}{a^2-a_c^2}\ln \Xi_{\rm EB}\,, \qquad 
E=\frac{q(e^{2\beta_\pm U}-1)}{[1+q^2(a^2-a_c^2)]\Xi_{\rm EB}}\,, 
\end{align}
\end{subequations}
where
\begin{align}
\label{}
\Xi_{\rm EB}\equiv  \frac{1+q^2(a^2-a_c^2)e^{2\beta_\pm U}}{1+q^2(a^2-a_c^2)} \,, \qquad 
\beta_\pm \equiv \pm\frac{a}{a_c}\sqrt{1+\beta^2}- \beta  \,. 
\end{align}
This solution has been derived in~\cite{ggs2009} for $a=0$ with $n=4$ and in~\cite{Maeda:2016ddh} for $a=0$ with arbitrary $n(\ge 4)$. 
We can restrict our analysis to the one particular branch in (\ref{dilaton-EB}), 
since two branches are interchanged by the simultaneous sign flip $\beta \to -\beta$ and $M\to -M$.
However, it is better suited for the present analysis to restrict to the $M>0$ case and consider two branches.
Expansion around $r\to \infty$,  the ADM mass and the charge are teased out as 
\begin{align}
\label{}
\ma M=\frac{(n-2)\Omega_{n-2}}{2\kappa_n}M \left(\beta -\frac{2a_c^2q^2\beta_\pm}{1+(a^2-a_c^2)q^2}\right)\,, \qquad 
\ma Q=\frac{2(n-3)\Omega_{n-2}Mq\beta_\pm }{\kappa_n[1+(a^2-a_c^2)q^2]}\,. 
\end{align}

Let us first tentatively suppose $g_{tt}<0$ for $r> 0$, i.e., assume 
$\Xi_{\rm EB}$ is strictly positive. 
Both branches of the metric admit a coordinate singularity at $r=0$. 
As $r\to 0$, we have $U\to \pi/2$, so that $\Xi_{\rm EB}$, $\varphi$ 
and $E$ remain finite there. 
We verify that the $r=0$ surface does not correspond to the 
scalar curvature singularity nor the p.p curvature singularity.
Inspection of (\ref{affine}) implies that  the affine parameter to the $r=0$ surface 
along the radial null geodesics is finite. 
One can then extend the spacetime across $r=0$  
by the replacement of $U(r)\to \pi/2-\arctan(2r^{n-3}/M)$ and $r^{n-3}=x$. 
In terms of this coordinate, 
the solution is rephrased as
\begin{align}
\label{dilaton-EBx}
\D s^2 =&-e^{-2\beta U_x}\Xi_x^{2a_c^2/(a^2-a_c^2)} \D t^2\notag \\
&+  \left[e^{2\beta U_x}
\Xi_x^{-2a_c^2/(a^2-a_c^2)}\right]^{1/(n-3)}\left(\frac{\D x^2}{(n-3)^2V_x^{(n-4)/(n-3)}}
+V_x^{1/(n-3)}\D \Omega_{n-2}^2\right)\,, \\
\varphi=&\pm\frac{\sqrt{1+\beta^2}}{a_c} U_x-\frac{a}{a^2-a_c^2}\ln \Xi_{x}\,, \qquad 
E=\frac{q(e^{2\beta_\pm U_x}-1)}{[1+q^2(a^2-a_c^2)]\Xi_{x}}\,, 
\end{align}
where 
\begin{align}
\label{}
V_x\equiv x^2+\frac{M^2}4\,, \qquad 
U_x\equiv\frac{\pi}2 -\arctan\left(\frac{2x}{M}\right)\,, \qquad 
\Xi_x\equiv \frac{1+q^2(a^2-a_c^2)e^{2\beta_\pm U_x}}{1+q^2(a^2-a_c^2)}\,.
\end{align}
Then, every component of the metric and its inverse is smooth at $x=0$, 
allowing one to extend into the $x<0$ region. The $x<0$ region corresponds respectively to 
the $r<0$ region in even dimensions and to the 
range of complex $r$ in odd dimensions. Defining $\ti x=-x$ and expanding 
the metric around $\ti x= \infty$, one gets
\begin{align}
\label{}
\D s^2\simeq & -e^{-2\pi \beta}\Xi_0^{2a_c^2/(a^2-a_c^2)} \left(1-\frac{M'}{\ti x}\right)\D t^2+
e^{2\pi \beta/(n-3)}\Xi_0^{-2a_c^2/[(n-3)(a^2-a_c^2)]} \left(1+\frac{M'}{(n-3)\ti x}\right)\notag \\
&\times
\left(\frac{\D \ti x^2}{(n-3)^2 \ti x^{2(n-4)/(n-3)}}+\ti x^{2/(n-3)}\D \Omega_{n-2}^2\right)\,, 
\end{align}
where 
\begin{align}
\label{}
\Xi_0\equiv \frac{1+e^{2\pi \beta_\pm}q^2(a^2-a_c^2)}{1+q^2(a^2-a_c^2)}
\,, \qquad 
M'\equiv -M \left(\beta-\frac{2a_c^2 q^2 e^{2\pi \beta_\pm}\beta_\pm}{1+e^{2\pi \beta_\pm}q^2(a^2-a_c^2)}\right)\,.
\end{align}
By a further replacement {$\ti x=e^{-\pi \beta}\Xi_0^{a_c^2/[(a^2-a_c^2)]}\ti r^{n-3}$} and 
$t=e^{\pi \beta}\Xi_0^{-a_c^2/(a^2-a_c^2)} \ti t$, the solution becomes asymptotically flat form 
\begin{align}
\label{}
\D s^2 \simeq -\left(1-\frac{2\kappa_n \ma M_{x<0}}{(n-2)\Omega_{n-2} \ti r^{n-3}}\right)\D \ti t^2
+\left(1+\frac{2\kappa_n \ma M_{x<0}}{(n-3)(n-2)\Omega_{n-2} \ti r^{n-3}}\right)\left(\D \ti r^2+\ti r^2 \D \Omega_{n-2}^2\right)\,,
\end{align}
with the ADM mass
\begin{align}
\label{}
\ma M_{x<0}= -\frac{(n-2)\Omega_{n-2}}{2\kappa_n} e^{\pi \beta}\Xi_0^{-a_c^2/(a^2-a_c^2)}
M \left(\beta-\frac{2a_c^2 q^2 e^{2\pi \beta_\pm}\beta_\pm}{1+e^{2\pi \beta_\pm}q^2(a^2-a_c^2)}\right)
 \,. 
\end{align}
In an analogous fashion, the asymptotic expansion of the electrostatic potential  in the $x\to -\infty$ limit
gives the electric charge as
\begin{align}
\label{}
\ma Q_{x<0} = \frac{2(n-3)\Omega_{n-2}Mq\beta_\pm
e^{2\pi (\beta_\pm+\beta)}\Xi_0^{-2a^2/(a^2-a_c^2)}}{\kappa_n[1+q^2(a^2-a_c^2)]}\,.
\end{align}
In the uncharged limit $q\to 0$, each ADM mass in the two regions necessarily has
the opposite sign unless $\beta=0$. For  the $q\ne 0$ case, the sign of ADM mass is not always flipped. 

\begin{figure}[t]
\begin{center}
\includegraphics[width=5cm]{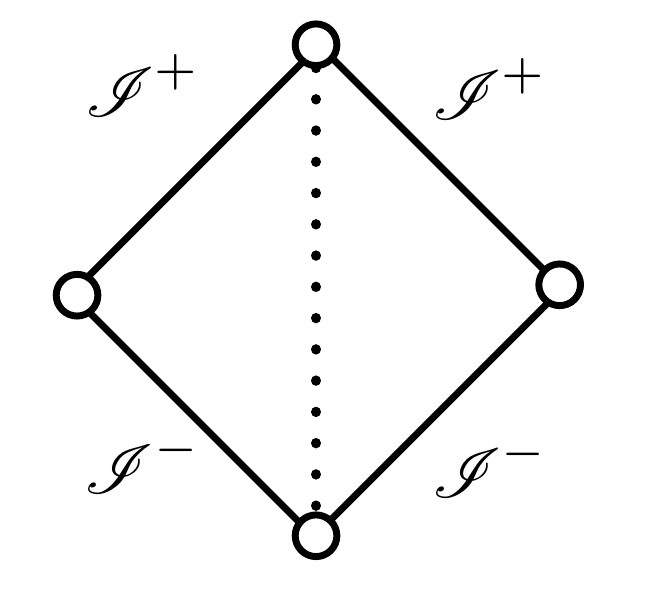}
\caption{A possible conformal diagram for the charged Ellis-Bronnikov solution describing a wormhole. 
In the noncritical case, 
the parameter should violate the conditions (\ref{EBnon:para}) to ensure the regularity of the solution. 
The dotted line corresponds to the $r=0$ surface, which is generically unequal to the locus of the
throat.}
\label{fig:EB}
\end{center}
\end{figure}

It turns out that the solution describes a regular charged wormhole which 
bridges the two asymptotically flat regions. It is stressed that the maximally extended spacetime
does not have a reflection symmetry around $x=0$.  
The global causal structure is shown in figure \ref{fig:EB}.

A commonly accepted notion which gives a tunneling surface of two asymptotically flat regions 
is the ``flaring-out'' wormhole throat \cite{Morris:1988cz,Kim:2013tsa}.  The locus $x=x_{\rm th}$ of the throat corresponds to the 
critical point of the areal radius 
$S=(e^{\beta U_x}V_x^{1/2}\Xi_x^{-a_c^2/(a^2-a_c^2)})^{1/(n-3)}$ and given by the 
solution to the following equation
\begin{align}
\label{throat}
%\left(1+q^2(a^2-a_c^2)e^{2\beta_\pm U_x(x_{\rm th})}\right)\left[
2x_{\rm th}-M\beta +q^2 e^{2\beta _\pm U_x(x_{\rm th})}[2(a^2-a_c^2)x_{\rm th}-a^2M\beta+a_c^2M
(\beta+2\beta_\pm)]
=0\,.
%\right]=0 \,. 
\end{align}
In the light of spacetime extension of the metric (\ref{dilaton-EB}), the 
natural ``entrance to the another universe'' is $r=0$, which is the boundary of the metric (\ref{dilaton-EB})
corresponding to the coordinate singularity. Generically this does not accord with the position of the throat (\ref{throat}) for the maximally extended spacetime. What is the most important to us is 
the fact that two asymptotically flat universes are joined, wherever the tunnel exists.

We have hitherto ignored the possibility that the solution admits a surface $\Xi_{x}=0$. 
The equation $\Xi_{x}=0$ is solved at the root $x\in \mathbb R$ of 
$1+q^2(a^2-a_c^2)e^{2\beta_\pm [\pi/2-\arctan(2x/M)]}=0$, which 
exists for 
\begin{align}
\label{EBnon:para}
(0<)a<a_c \quad {\rm and} \quad 
\left\{
\begin{array}{cc}
 1<(a_c^2-a^2)q^2 < e^{-2\pi \beta_\pm}\,,     &( {\rm for}~\beta_\pm<0)   \\
e^{-2\pi \beta_\pm} <(a_c^2-a^2)q^2 < 1\,,    & ({\rm for}~\beta_\pm>0)
\end{array}
\right. \,.
\end{align}
If these conditions are fulfilled, 
the solution corresponds to the spacetime admitting a singular surface at $\Xi_{\rm EB}=0$. 
Thus, the regular wormhole configuration should not satisfy the conditions (\ref{EBnon:para}). 
We shall not attempt to go any further into this singular case.

%======================================%
%<<<<<<<<<<<< SECTION II >>>>>>>>>>>>>>%
%======================================%
\section{Critical case}
\label{sec:critical}

In this section, we present solution-generating transformations in the system (\ref{action-dil2}) for the special case with a phantom dilaton field ($\epsilon=-1$) and the critical coupling $a=a_{c}=\sqrt{2(n-3)/(n-2)}$. In this case, one sees that the variables introduced in (\ref{phivar}) are ill-defined. This asks for a separate analysis. 

\subsection{Solution-generating transformations}

With the critical coupling $a=a_{c}$ for $\epsilon=-1$, the target space metric (\ref{targetmet}) reduces to
\begin{align}
\label{targetspace-critical}
\D s_T^2=2 \D u \D v-2e^{-2a_c u} \D E^2 \,, 
\end{align}
where 
\begin{align}
\label{}
u\equiv \varphi-\sigma\,, \qquad v\equiv -(\varphi+\sigma)\,.
\end{align} 
This metric describes a symmetric pp-wave admitting a 
covariantly constant Ricci tensor $\mathcal D_a \mathcal R_{bc}=0$ 
and a covariantly constant null Killing vector $\partial/\partial v$. 
As shown by (B.31) in~\cite{Nozawa:2019dwu}, there exist four Killing vectors $\xi_{(A)}~(A=1,2,3,4)$ in the target space (\ref{targetspace-critical}), which are given by\footnote{
Observe that the Killing vectors in (\ref{KVsnoncrit}) are well-defined in the critical limit ($a\to a_c$ with 
$\epsilon=-1$). The principal difference comes from the algebra they constitute, which 
alters considerably the finite transformations. 
}
\begin{align}
\label{KVscrit}
\xi_{(1)} =\partial_v \,, \qquad 
\xi_{(2)} =\frac 1{a_c}\partial_u+E \partial_E \,, \qquad
\xi_{(3)}=\partial_E \,, \qquad 
\xi_{(4)} =E\partial_v+\frac{e^{2a_c u}}{4a_c}\partial_E \,, 
\end{align}
forming the closed four-dimensional algebra with the following nonvanishing
commutators\footnote{This is nothing but the Heisenberg algebra appeared in the 
Wess-Zumino-Witten model \cite{Nappi:1993ie}. 
By identifying 
$J=-\xi_{(2)}$, $P_0=(\xi_{(3)}-\xi_{(4)})/\sqrt 2$, $P_1=(\xi_{(3)}+\xi_{(4)})/\sqrt 2$, $T=\xi_{(1)}$ in the notation of \cite{Nappi:1993ie}, 
the algebra (\ref{critalg}) is summarized as $[J, P_i]=\epsilon_{ij}P^j$, $[P_i, P_j]=\epsilon_{ij}T$, 
where indices are raised and lowered by $\eta_{ij}={\rm diag}(-1,1)$. 
This is the central extension of the two-dimensional Poincar\'e algebra with 
a central charge $T$.}
\begin{align}
\label{critalg}
[\xi_{(2)},\xi_{(3)}]=-\xi_{(3)} \,, \qquad 
[\xi_{(2)},\xi_{(4)}]=\xi_{(4)}\,, \qquad 
[\xi_{(3)},\xi_{(4)}]=\xi_{(1)}\,.
\end{align}
It turns out that $\xi_{(1)}$ is the center of the algebra and the Killing-Cartan metric for the above algebra is 
degenerate, so that the algebra is non-semisimple. It follows that the symmetry of the nonlinear sigma model
changes substantially at this critical coupling case. 

The finite transformations $\{u,v,E\}\to \{u',v',E'\}$ corresponding to the Killing vectors $\xi_{(1)}$--$\xi_{(4)}$ read respectively
\begin{subequations}
\begin{align}
{\rm [I_c]}~~~~& u'=u, \qquad v'=v+c_1 \,, \qquad E'=E,\label{crit:tr1}\\
{\rm [II_c]}~~~~& u'=u+c_2, \qquad v'=v \,, \qquad E'=E e^{a_c c_2 }\,, \label{crit:tr2}\\
{\rm [III_c]}~~~~& u'=u, \qquad v'=v \,, \qquad E'=E+c_3 \,, \label{crit:tr3}\\
{\rm [IV_c]}~~~~& u'=u, \qquad v'=v+E c_4 +\frac{1}{8a_c}e^{2a_cu}c_4^2 \,, \qquad E'=E+\frac 1{4a_c}e^{2a_cu}c_4\,,\label{crit:tr4}
\end{align}
\end{subequations}
where $c_1$--$c_4$ are constants. 
Transformations I${}_{c}$--III${}_{c}$ are pure gauge, while 
the transformation IV${}_{c}$ gives rise to the Harrison transformation \cite{Harrison}
which allows one to charge up the neutral solution.

Let us now apply the transformations I${}_{c}$--IV${}_{c}$ given by (\ref{crit:tr1})--(\ref{crit:tr4}) to generate new exact solutions.
We assume that the seed solution $\{\sigma, \varphi, E\}=\{\sigma_0, \varphi_0, E_0\}$ is neutral 
$(E_0=0)$ and fulfill the fall-off condition ($\sigma_0 \to 0$ and $\varphi_0\to 0$ as $r\to \infty$), 
corresponding to the preservation of asymptotic flatness. 
An elementary calculation yields that the new solution $\{\sigma, \varphi, E\}$ preserving asymptotic flatness is given by 
\begin{subequations}
\label{chargetr:crit}
\begin{align}
\sigma&=\sigma_0 -\frac{q^2}{a_c}\left(e^{2a_c(\varphi_0-\sigma_0)}-1\right)\,, \\
\varphi&=\varphi_0-\frac{q^2}{a_c} \left(e^{2a_c(\varphi_0-\sigma_0)}-1\right)\,, \\
E&=\frac{q}{a_c}\left(e^{2a_c(\varphi_0-\sigma_0)}-1\right)\,.
\end{align}
\end{subequations}
Here $q$ is a parameter corresponding to the electric charge.
Using the formulae (\ref{chargetr:crit}), we will construct exact static and asymptotically flat charged solutions in the phantom case ($\epsilon=-1$) with the critical coupling $a=a_{c}$.

\subsection{Charged Fisher solution}

We first take the $n(\ge 4)$-dimensional phantom Fisher solution (\ref{FJNW}) as a seed, for which the functions $\sigma=\sigma_0$ and $g_{IJ}$ in the metric (\ref{metric}) and the scalar field $\varphi=\varphi_0$ are given by (\ref{JNWseed}).
Applying the transformation (\ref{chargetr:crit}) to this solution, we obtain the charged Fisher solution for the critical coupling as
\begin{subequations}
\label{chargedFisher:crit}
\begin{align}
\label{}
\D s^2&=-f(r)^{\alpha}e^{2q^2h(r)}\D t^2
+f(r)^{-(n-4+\alpha)/(n-3)} e^{-2q^2h(r)/(n-3)}\left(\D r^2+r^2 f(r)\D \Omega_{n-2}^2\right) \,,
\\
\varphi &=\pm \frac{\sqrt{{\alpha^2-1}}}{2a_c}\ln f(r)-\frac{q^2}{a_c}h(r)\,, \qquad 
E=\frac{q}{a_c} h(r)\,. 
\end{align}
\end{subequations}
where $f(r)= 1-M/r^{n-3}$ and 
\begin{align}
\label{}
h(r)\equiv  f(r)^{\pm\sqrt{{\alpha^2-1}}+\alpha}-1\,. 
\end{align}

\subsubsection{$\alpha^2=1$ case}

Let us first discuss the $\alpha^2=1$ case. 
Since the $\alpha =+1$ and $\alpha=-1$ cases are interchanged by $M\to -M$, we can 
consider only for the $\alpha=+1$ case, for which the solution reads
\begin{subequations}
\label{SchGibbons}
\begin{align}
\D s^2_{\alpha=1}=& -f(r)e^{-2q^2M/r^{n-3}} \D t^2+e^{2q^2M/[(n-3)r^{n-3}]}
\left(\frac{\D r^2}{f(r)}+r^2 \D \Omega_{n-2}^2\right)\,,  \\
\varphi=& \frac{q^2M}{a_cr^{n-3}}\,, \qquad 
E=-\frac{qM}{a_c r^{n-3} } \,,
\end{align}
\end{subequations}
where $f(r)=1-M/r^{n-3}$. 
It is interesting to observe that this solution interpolates the Schwarzschild solution and the 
Gibbons solution. Indeed, the latter solution can be recovered by setting $M'=2q^2M$, $q'=qM$  and $q'\to 0$ with $M'$ kept finite.
The corresponding four dimensional solution was obtained in~\cite{Gibbons:1996pd,Gao:2006iw}. 

The Ricci scalar and the p.p frame component of the Ricci tensor are given by
\begin{align}
\label{}
R=& -\frac{(n-3)M^2q^2}{r^{2(n-2)}}e^{-2Mq^2/[(n-3)r^{n-3}]} \left[n-4+(n-2)q^2
\left(1-\frac{M}{r^{n-3}}\right)
\right]\,, \\
R_{\mu\nu}k^\mu k^\nu=& -\frac{(n-3)(n-2)M^2 q^4}{r^{2(n-2)}}e^{2(n-4)Mq^2/[(n-3)r^{n-3}]} \,. 
\end{align}
It follows that $r=0$ is a scalar curvature singularity for $M<0$ and a p.p curvature singularity for 
$M>0$. 
The solution (\ref{SchGibbons}) then describes the asymptotically flat black hole with a regular event horizon at $r=r_+\equiv M^{1/(n-3)}$ for $M>0$ and a naked singularity at $r=0$ for $M<0$. 
Using (\ref{affine}) and noting $\int ^r r^b e^{c/r^{n-3}}\D r$ diverges as $r\to 0$ for $c>0$, 
the affine parameter to $r=0$ is finite for $n=4$ and $n\ge 5$ with $M>0$, whereas it 
is infinite for $n\ge 5$ with $M<0$. Inspecting (\ref{tortoise}), the $r=0$ is null for 
$M>0$ and timelike for $M<0$. In the interior of the black hole, the proper time 
along the timelike geodesics is infinite for $M>0$. 
The causal structure of the solution with $M>0$ is therefore (III) in figure \ref{fig:PDFisher}.

The ADM mass and the electric charge are given by
\begin{align}
\label{}
\ma M=\frac{(n-2)\Omega_{n-2}}{2\kappa_n}M(1+2q^2)\,, \qquad 
\ma Q=\frac{2(n-3)\Omega_{n-2}}{a_c\kappa_n} Mq \,. 
\end{align}
For $M>0$, 
the surface gravity, the area and the electrostatic potential at the event horizon
$r=M^{1/(n-3)}$  are given respectively by 
\begin{subequations}
\begin{align}
\label{}
\kappa=&\frac 12 (n-3)M^{-1/(n-3)} e^{-(n-2)q^2/(n-3)} \,, \\
A_{\rm H}=&\Omega_{n-2} M^{(n-2)/(n-3)} e^{(n-2)q^2/(n-3)}\,, \\
\Phi_{\rm H}=&\frac{q}{a_c} \,, 
\end{align}
\end{subequations}
satisfying the first law (\ref{1st}). 

\subsubsection{$\alpha^2\ne 1$ case}

Next we shall consider the $\alpha^2\ne 1$ case. Then, 
the plus-minus branches in (\ref{chargedFisher:crit}) are interchanged by $M\to -M$, $\alpha \to -\alpha$. We then deal with the plus branch for all range of $M$ and $\alpha$ in the remainder. 
This gives (i) $M>0$ with $\alpha > 1$, (ii) $M>0$ with $\alpha< -1$, 
(iii) $M<0$ with $\alpha > 1$ and (iv) $M<0$ with $\alpha <-1$.
For $M>0$, the range of $r$ is $r>r_s \equiv M^{1/(n-3)}$, while 
for  $M<0$, the range of $r$ is $r>0$.
The Ricci scalar and the p.p component of the Ricci tensor read respectively
\begin{align}
\label{}
R=&\frac{(n-3)M^2}{r^{2(n-2)}}e^{2q^2 h/(n-3)} f^{(2-n+\alpha)/(n-3)}
\Bigl\{
(n-2)q^4\left[1-2\alpha(\alpha+\sqrt{\alpha^2-1})\right]f^{2(\alpha+\sqrt{\alpha^2-1})}
\notag \\&
-q^2\left[2+(n-6)\alpha(\alpha+\sqrt{\alpha^2-1})\right]f^{\alpha+\sqrt{\alpha^2-1}}
+\frac 14(n-2)(1-\alpha^2)
\Bigr\}\,, 
\end{align}
and 
\begin{align}
\label{}
R_{\mu\nu}k^\mu k^\nu =&-\frac{(n-3)(n-2)M^2}{r^{2(n-2)}}f(r)^{-\frac{(n-4)\alpha+(n-2)}{n-3}}e^{-\frac{2(n-4)}{n-3}q^2 h(r)} 
\notag \\ & \times \left(\frac 12 \sqrt{\alpha^2-1}-(\alpha+\sqrt{\alpha^2-1})q^2 f(r)^{\alpha+\sqrt{\alpha^2-1}}\right)^2 \,. 
\end{align}
In case (i) (case (iv)), 
$h(r) \to -1$ as $r\to r_s$ ($r\to 0$), in which the electric charge does not affect the structure of 
$r=r_s$ ($r=0$) surface compared to the neutral one, so that these surfaces are either scalar curvature/p.p curvature singularities. 
In case (ii) (case (iii)), $h(r) \to +\infty$ as $r\to r_s$ ($r\to 0$), so that these surfaces are scalar curvature singularities. It follows that the $\alpha^2\ne 1$ solution does not contain regular configurations.

\subsection{Charged Gibbons solution}

Next we employ the $n(\ge 4)$-dimensional Gibbons solution (\ref{Gibbons}) as a seed solution, for which the functions $\sigma=\sigma_0$ and $g_{IJ}$ in the metric (\ref{metric}) and the scalar field $\varphi=\varphi_0$ are given by (\ref{seed:Gibbons}).
Then we have $2a_c(\varphi_0-\sigma_0)=\pm H-H$, so that only the minus branch in (\ref{seed:Gibbons}) generates a nontrivial solution. 
Applying the transformation (\ref{chargetr:crit}) to this solution, we obtain the charged Gibbons solution for the critical coupling as
\begin{subequations}
\begin{align}
\label{critchargedsol}
\D s^2&=- e^{-\hat H}\D t^2+e^{\hat H/(n-3)} 
h_{IJ}\D  y^I \D y^J \,,  \\
\varphi&=-\frac 1{2a_c}\left[H+2q^2(e^{-2H}-1)\right]\,, \qquad 
E= \frac{q}{a_c} (e^{-2H}-1) \,,
\end{align}
\end{subequations}
where $H$ is a harmonic function satisfying $\Delta_h H=0$ and 
\begin{align}
\hat H\equiv  H-2q^2(e^{-2H}-1)\,.
\end{align}

To concrete, let us consider the spherically symmetric case 
$H=M/r^{n-3}$.
Since we have $\hat H\sim M/r^{n-3}\to \infty $ as $r\to 0$ for $M>0$ and 
$\hat H\sim -2q^2 e^{-2M/r^{n-3}}\to- \infty $ as $r\to 0$ for $M<0$, 
the Ricci scalar and similarly all curvature invariants are divergent at $r=0$ for $M<0$. 
The  parallelly propagated frame component of the Ricci tensor reads 
$R_{\mu\nu}k^\mu k^\nu \propto -e^{(n-4)M/[(n-3)r^{n-3}]}(1-q^2 e^{-2M/r^{n-3}})/r^{2(n-2)}$ as $r\to 0$. 
It follows that the the spacetime is singular in any parameter region.

\subsection{Charged Ellis-Bronnikov solution}

Lastly, we take the $n(\ge 4)$-dimensional Ellis-Bronnikov solution (\ref{Ellis-Bronnikov}) as a seed solution, for which the functions $\sigma=\sigma_0$ and $g_{IJ}$ in the metric (\ref{metric}) and the scalar field $\varphi=\varphi_0$ are given by (\ref{seed:Ellis}).
Applying the transformation (\ref{chargetr:crit}) to this solution, we obtain 
\begin{subequations}
\begin{align}
\label{dilaton-EB-critical}
\D s^2=&-e^{-2\beta U(r)+2 q^2 W(r)} \D t^2+\left[e^{2\beta U(r)-2q^2 W(r)} V(r)\right]^{1/(n-3)}
\left(\frac{\D r^2}{V(r)}+r^2 \D \Omega_{n-2}^2 \right)\,, \\
\varphi=& \pm\sqrt{1+\beta^2} \frac{U(r)}{a_c}-\frac{q^2}{a_c}W(r)\,, \qquad E=\frac{q}{a_c}W(r) \,,
\end{align}
\end{subequations}
where $U(r)=\arctan(M/(2r^{n-3}))$, $V(r)= 1+M^2/(4r^{2(n-3)})$, and 
\begin{align}
W(r)\equiv e^{2\beta_{c, \pm}U(r)}-1 \,, \qquad 
\beta_{c, \pm}\equiv \pm\sqrt{1+\beta^2}-\beta\,.
\end{align}
This is a higher dimensional generalization of the one  obtained in~\cite{Goulart:2017iko}. 

From the asymptotic form of the metric and the field strength, one can read the ADM mass and the electric charge as
\begin{align}
\label{}
\ma M= \frac{(n-2)\Omega_{n-2}}{2\kappa_n}M \left(\beta-2q^2
\beta_{c, \pm} \right)\,, \qquad
\ma Q=\frac{2(n-3)\Omega_{n-2}}{a_c \kappa_n} Mq
\beta_{c, \pm}\,. 
\end{align}
Contrary to the noncritical case, 
there seems no restriction on $|\ma Q/\ma M|$ for the regularity of the solution.

Let us now pay attention to the $r=0$ surface. From (\ref{tortoise}) and (\ref{affine}), this surface is timelike and can be reached within a finite affine time for a radial null geodesics.
All the curvature invariants and the components of the Riemann tensor in a parallelly propagated frame
seem to be perfectly well behaved as $r\to 0+$. One can then extend the spacetime across the $r=0$ surface by replacing 
$U(r)=\pi/2-\arctan(2r^{n-3}/M)$ with $r^{n-3}=x$. 
It turns out that the solution describes a regular wormhole in the two-sided asymptotically flat regions. 
The Penrose diagram is the same as the noncritical case in figure \ref{fig:EB}.

Since the remaining calculations for conserved quantities are parallel with the non-critical case, we show only the final outcome. The ADM mass and the electric charge in the 
$x<0$ region read
\begin{align}
\label{}
\ma M_{x<0}&=-\frac{(n-2)\Omega_{n-2}}{2\kappa_n}M(\beta-2q^2 \beta_{c, \pm}e^{2\pi \beta_{c,\pm}})
e^{\pi \beta-q^2(e^{2\pi \beta_{c, \pm}}-1)} \,, \notag \\
\ma Q_{x<0}&=\frac{2(n-3)\Omega_{n-2}Mq\beta_{c,\pm}}{a_c \kappa_n} e^{2\pi(\beta+\beta_{c,\pm})-2q^2
(e^{2\pi \beta_{c,\pm}}-1)} \,.
\end{align}

This solution exemplifies the static wormhole solution in~\cite{Rogatko:2018crz}
(see also \cite{Yazadjiev:2017twg,Lazov:2017tjs,Rogatko:2018smj} for previous works), in which  
the uniqueness theorem of wormholes has been proven, but the desired solution has not been displayed explicitly.

%======================================%
%<<<<<<<<<<<< SECTION  >>>>>>>>>>>>>>%
%======================================%
\section{Summary and concluding remarks}
\label{sec:conclusion}

In the present paper, we have established a method to generate static charged solutions in the Einstein-Maxwell-(phantom-)dilaton system in arbitrary $n(\ge 4)$ dimensions, based on the symmetry of the target space for the nonlinear sigma model in the truncated action. This is a generalization of the results with a conventional dilaton field~\cite{Galtsov:1995mb,Yazadjiev:2005hr}, and the phantom-dilaton case in four dimensions~\cite{Gibbons:1996pd,AzregAinou:2011rj}. We pointed out that there exists a critical case of the coupling in the phantom case, for which the symmetry of the target space and therefore the derived solutions are different considerably in expressions. 

As applications of this scheme, we have constructed the dilatonic charged versions of the Fisher solution, the Gibbons solution, and the Ellis-Bronnikov solution, respectively. Inferring from the complexity of the solutions, these solutions would have been inaccessible without the solution-generating technique. 
We have endeavored to list the physical and causal properties of these solutions in detail. This is summarized as follows:
\begin{itemize}
  \item 
 The (non)critical charged Fisher solution parameterized by $\alpha $,  $M$ and $q$ admits a horizon only for the $\alpha^2=1$ case. We have exposed that the degenerate limit of the $\alpha^2=1$ noncritical solution is still singular due to the p.p curvature singularity. This type of singularity might not be so harmful, but deserves an abnormal nature of the spacetime property. The (non)critical  solutions with $\alpha^2\ne 1$ are always singular.
 
 \item
 The (non)critical charged Gibbons solution parameterized by $M$ and $q$ always admit a scalar or p.p curvature singularity at $r=0$, which is not covered by a horizon. We therefore have to abandon the picture that the solution bridges two different regular universes at the throat, since one side of the universe is necessarily  singular \cite{Boonserm:2018orb}. 
This highlights the importance of investigating the geodesic motion to clarify the global causal structure of the spacetime.

\item
We have shown that the dilatonic charged Ellis-Bronnikov solution, parameterized by $\beta$, $M$ and $q$, represents a regular wormhole connecting two asymptotically flat regions. 
In the noncritical case, the parameters are constrained not to satisfy (\ref{EBnon:para}).  
The analysis of global structure has been done only for $n=4$ with the critical coupling $a=a_{c}$ in~\cite{Goulart:2017iko}. 
In higher dimensions, we have presented for the first time the exact solution and its maximal extension.
The solution presented here (\ref{dilaton-EB-critical}) fills also a gap in~\cite{Rogatko:2018crz}, where the uniqueness of static spherically symmetric traversable wormholes with two asymptotically flat regions has been proven in the critical coupling case, but the corresponding unique wormhole solution has not been obtained in that paper. 

\end{itemize}

Let us conclude the present paper by stating some future extensions of our results.

\begin{itemize}
  \item 
For the $\alpha^2=1$ noncritical Fisher solution, we have a black hole horizon for $r_+>0$, whose 
degenerate limit $r_+\to r_-$ gives a singular surface. This property affects the 
evaporation process of a black hole in the non-phantom case \cite{Koga:1995bs}. This can be inspected from the behavior of the Hawking temperature (i.e., the surface gravity (\ref{surfacegrav}) with $\epsilon=1$)
\begin{align}
\label{}
T &\to \left\{
\begin{array}{cc}
 0  \,,   &   (a<\sqrt{n-3}a_c) \\
 +\infty \,,     &   (\sqrt{n-3}a_c<a )
\end{array}
\right.\,,
\end{align}
as $r_+\to r_-$. From table \ref{table:FisherBH}, 
the degenerate limit $r_+\to r_-$ exists only for $a<a_c$ in the phantom case. 
For the geometro-thermodynamics, a related discussion can be found in \cite{Quevedo:2016swn}. 
It is interesting to explore how the whole evaporation process 
depends on the coupling constant of the dilaton in the phantom case with $a>a_c$. 

\item
Except for the case with the critical coupling constant $a=a_c$ \cite{Rogatko:2018crz,Lazov:2017tjs}, the uniqueness of the charged wormhole solutions has not been explored yet. We believe that this can be shown along with the scheme based upon the divergence-type equations \cite{Nozawa:2018kfk}. 

\item
Although we have established a method to generate static solutions in this paper, a complete classification of static solutions is remained open. The rationale is two-fold. 
One reason is that the present solution-generating method retains the base space metric $g_{IJ}$ invariant. This places a certain restriction to the derived solutions. The other reason is that there might appear a solution which does not admit a neutral limit. The Nariai-type solutions are of this sort \cite{Nariai}. For this purpose, what is imperative is the brute-force integration of Einstein's equations, which was the road taken in a comparatively simple system~\cite{Clement:2009ai,Maeda:2016ddh}. Such a complete classification in the Einstein-Maxwell-(phantom-)dilaton system in arbitrary dimensions is challenging, but one of the directions of our future investigation. 

\item
Wormholes in asymptotically anti-de Sitter spacetimes have attracted much attention recently, 
in the context of holographic entanglement. 
The extension of the static solutions into asymptotically (anti-)de Sitter spacetimes will be reported in the  forthcoming paper \cite{paperIII}. 

\end{itemize}

\subsection*{Acknowledgements}

M. N. thanks  Hideki Maeda and Cristi{\'a}n Mart\'{\i}nez for discussions. 
This work  is partially supported by 
Grant-in-Aid for Scientific Research (A) from JSPS 17H01091 and (C) 20K03929.

\appendix
\renewcommand{\theequation}{A.\arabic{equation}}
\setcounter{equation}{0}

\section{Curvatures}
\label{sec:curv}

In the body of text, we consider the $n$-dimensional spacetime to 
admit ``pseudo-spherical'' symmetry. This appendix summarizes some technical results toward the physical discussion. 
The line element is supposed to be given by
%------------------- metric ansatz ---------------------%
\begin{align}
g_{\mu \nu }\D x^\mu \D x^\nu =g_{AB}(y)\D y^A\D y^B +S^2(y) \gamma _{ij}(z)\D z^i\D z^j,
\label{eq:ansatz}
\end{align} 
where
$A,B = 0, 1;~i,j = 2, ..., n-1$. 
Here $S$ is a scalar on the two-dimensional Lorentzian spacetime $(M^2, g_{AB})$ and $\gamma_{ij}$ is the metric on the maximally symmetric space $(K^{n-2}, \gamma _{ij})$ with its sectional curvature $k = \pm 1, 0$. 
The non-vanishing components of the Levi-Civita connections are
\begin{align}
\begin{aligned}
{\Gamma ^A}_{BC}&={}^{(2)}{\Gamma ^A}_{BC }(y),\quad 
{\Gamma ^i}_{jk}={\hat{\Gamma} ^i}_{~jk}(z), \\
{\Gamma ^A}_{jk}&=-S(D^A S) \gamma _{jk},\quad 
{\Gamma ^i}_{jA}=\frac{D_A S}{S}{\delta ^i}_j, 
\end{aligned}
\end{align}
where the superscript (2) stands for the two-dimensional quantity and $D_A$ is the covariant 
derivative of  $g_{AB}$.
The non-vanishing components of the Riemann tensors are
\begin{subequations}
\label{eq:Riemann}
\begin{align}
{{R}^A}_{BCD}&={}^{(2)}{{R}^A}_{BCD},\\
{{R}^A}_{iBj}&=-S(D^A D_B S)\gamma _{ij},
\\
{{R}^i}_{jkl}&=[k-(DS)^2]({\delta ^i}_k\gamma _{jl}-{\delta ^i}_l\gamma _{jk})\,, 
\end{align}
\end{subequations}
where $(DS)^2\equiv 
(D_AS)(D^AS)$. 
The Ricci tensor and the Ricci scalar read
\begin{subequations}
\label{eq:Ricci}
\begin{align}
{R}_{AB}&={}^{(2)}{R}_{AB}-(n-2)\frac{D_AD_BS}{S},\\
{R}_{ij}&=\left\{-S D^2S+(n-3)[k-(DS)^2]\right\}\gamma _{ij},  \\
{R}&={}^{(2)}{R}-2(n-2)\frac{D^2S}{S}+(n-2)(n-3)\frac{k-(DS)^2}{S^2}, 
\end{align}
\end{subequations}
where $D^2S\equiv D_AD^AS$. 
The Kretschmann scalar $\mathcal K\equiv {R}^{\mu\nu\rho\sigma}{R}_{\mu\nu\rho\sigma}$ is 
\begin{align}
\mathcal K=
&{}^{(2)}{R}^2 
+4(n-2)\frac{(D_AD_BS)(D^AD^BS)}{S^2}+2(n-2)(n-3)\frac{(k-(DS)^2)^2}{S^4}\,. 
\label{Kre}
\end{align}

We are interested in the static spacetime, for which the metric takes the form
\begin{align}
\label{}
\D s^2=-f_1(r) \D t^2+f_2 (r ) \D r^2 +S^2(r) \D \Sigma_{k,n-2}^2 \,. 
\end{align}
To extract the causal nature of the $r={\rm const.}$ surface, it is advantageous to take 
the two-dimensional part to be conformally flat form 
\begin{align}
\label{tortoise}
\D s_2^2=-f_1(r)\D t^2+f_2(r)\D r^2=-f_1(r(r_*)) (\D t^2-\D r_*^2) \,, \qquad 
r_* =\int ^r \sqrt{\frac{f_2(r)}{f_1(r)}}\D r \,.
\end{align}
If the tortoise coordinate $r_*$ diverges at some $r$, it corresponds to the null surface. 
If $r_*$ remains finite there and $f_1(r(r_*))>0$ ($f_1(r(r_*))<0$), 
it corresponds to the timelike (spacelike) surface. 

To explore the causal structure of the spacetime, 
it is important to keep track of the behavior of the radial null geodesics. 
The tangent vector of a radial null geodesic
$k^\nu\nabla_\nu k^\mu=0$ is given by
\begin{align}
\label{}
k^\mu =f_1^{-1} (\partial_t)^\mu\pm (f_1f_2)^{-1/2}(\partial_r)^\mu\,. 
\end{align} 
Accordingly, the affine parameter $\lambda $ along the null geodesics reads 
\begin{align}
\label{affine}
\pm \lambda =\int ^r \sqrt{f_1(r)f_2(r)}\D r \,. 
\end{align}
The Ricci tensor component is 
\begin{align}
\label{Rkk}
R_{\mu\nu}k^\mu k^\nu =(n-2)\frac{(f_1f_2'+f_1'f_2)S'-2f_1f_2 S''}{2f_1^2f_2^2S}\,.
\end{align}
As shown in \cite{paperI}, the divergence of this component implies the existence of a 
p.p curvature singularity. To see this, 
let $e^{\hat i}{}_i$ be an orthonormal frame with 
$\gamma_{ij}=\delta_{\hat i\hat j}e^{\hat i}{}_ie^{\hat j}{}_j$ and we define
$ E^{\hat i}{}_\mu = S(r) e^{\hat i}{}_i (\D z^i)_\mu$. Defining 
\begin{align}
\label{}
n^\mu=\frac 12 (\partial_t)^\mu \mp \frac{\sqrt{f_1}}{2\sqrt{f_2}}(\partial_r)^\mu \,, \qquad
n_\mu n^\mu=0\,,
\end{align}
we have $g_{\mu\nu}=-2k_{(\mu}n_{\nu)}+\delta_{\hat i\hat j}E^{\hat i}{}_\mu E^{\hat j}{}_{\nu}$ with
\begin{align}
\label{}
k^\nu \nabla_\nu n^\mu= k^\nu \nabla_\nu E_{\hat i}{}^\mu =0 \,.
\end{align}
Specifically, $\{k^\mu, n^\mu, E_{\hat i}{}^\mu\}$ constitutes the frame which is 
parallelly propagated along $k^\mu$. 
Since we have
$R_{\mu \nu\rho\sigma}k^\mu E_{\hat i}{}^\nu k^\rho E_{\hat j}{}^\sigma=(n-2)^{-1}R_{\mu\nu}k^\mu k^\nu\delta_{\hat i\hat j}$, the divergence of $R_{\mu\nu}k^\mu k^\nu$ is tantamount to the p.p curvature singularity.

\end{document}